\pdfoutput=1
\documentclass[aps,prc,twocolumn,floatfix,superscriptaddress,nofootinbib,preprintnumbers,longbibliography,amsmath,amsthm,amssymb]{revtex4-2}
\usepackage{graphicx}

\usepackage{amsfonts}
\usepackage{color}
\usepackage[colorlinks=true,linkcolor=blue,citecolor=blue]{hyperref}
\usepackage{dcolumn}
\usepackage{bm}

\newcommand{\be}{\begin{equation}}
\newcommand{\ee}{\end{equation}}
\newcommand{\dd}{\mathrm{d}}
\newcommand{\br}{\boldsymbol{r}}

\begin{document}
\allowdisplaybreaks[1]

\title{Isovector giant monopole and quadrupole resonances in a Skyrme energy density functional approach with axial symmetry
}

\author{Kenichi Yoshida}
\email[E-mail: ]{kyoshida@ruby.scphys.kyoto-u.ac.jp}
\affiliation{Department of Physics, Kyoto University, Kyoto, 606-8502, Japan}

\preprint{KUNS-2884}
\date{\today}

\begin{abstract}
\begin{description}
\item[Background] 
Giant resonance (GR) is a typical collective mode of vibration. 
The deformation splitting of the isovector (IV) giant dipole resonance is well established. 
However, the splitting of GRs with other multipolarities is not well understood. 
\item[Purpose] I explore the IV monopole and quadrupole excitations 
and attempt to obtain the generic features of IV giant resonances in deformed nuclei by investigating the neutral 
and charge-exchange channels simultaneously. 
\item[Method] I employ a nuclear energy-density functional (EDF) method: the Skyrme--Kohn--Sham--Bogoliubov and the quasiparticle 
random-phase approximation are used to describe the ground state and the transition to excited states.
\item[Results] I find the concentration of the monopole strengths in the energy region of the isobaric analog or Gamow--Teller resonance 
irrespective of nuclear deformation, and the appearance of a high-energy giant resonance 
composed of the particle--hole configurations of $2\hbar \omega_0$ excitation. 
Splitting of the distribution of the strength occurs in the giant monopole and quadrupole resonances due to deformation. 
The lower $K$ states of quadrupole resonances appear lower in energy and possess the enhanced strengths in the prolate configuration, 
and vice versa in the oblate configuration, 
while the energy ordering depending on $K$ is not clear for the $J=1$ and $J=2$ spin-quadrupole resonances.
\item[Conclusions]
The deformation splitting occurs generously in the giant monopole and quadrupole resonances. 
The $K$-dependence of the quadrupole transition strengths is largely understood by the anisotropy of density distribution. 
\end{description}
\end{abstract}

\maketitle

\section{Introduction}\label{intro}

The response of a nucleus to an external field induces various modes of excitation, 
reflecting many-nucleon correlations and inter-nucleon interactions in the nuclear medium. 
Since the external fields are classified by quantum numbers, 
the collective modes of motion are selectively excited~\cite{har01}; 
the nuclear response is characterized by the transferred angular momentum $\Delta L$, 
spin $\Delta S$, isospin $\Delta T$, and particle number $\Delta N$.

The isovector (IV) giant dipole resonance (GDR) represented as $\Delta L=1, \Delta S=0, \Delta T=1, \Delta N=0$
is one of the well studied collective vibrational modes of excitation 
among various types of giant resonance (GR)~\cite{ber75}. 
The GDR is an oscillation of protons against neutrons represented as $\Delta T_z=0$ 
and can be seen in a wider perspective 
when it is considered as a single component $\Delta T_z=0$ of the IV dipole modes~\cite{BM2,aue83,izu83,aue84}. 
The additional components $\Delta T_z= \pm 1$ represent the charge-exchange modes. 
In addition to the Coulomb potential, with the presence of excess neutrons, {\it i.e.}, deformation in isospin space, 
the IV strengths reveal the splitting for $\Delta T_z=0, \pm1$~\cite{BM2}. 
The charge-exchange excitations have attracted interest  
not only because they reflect the isospin and spin--isospin character of a nucleus 
but because they have a relevance for nuclear $\beta$-decay, thus connecting strong and weak interactions~\cite{ost92}. 
However, there has been little study of the giant multipole resonances other than the dipole, 
isobaric analog (IAR), and Gamow--Teller (GTR) resonances~\cite{har01}.  

Extensive theoretical works in Refs.~\cite{aue83,izu83,aue84} 
opened up an avenue of the study for the IV multipole excitations other than $\Delta L=1$. 
Recent experimental progress has enabled precise measurements of the electric quadrupole resonance~\cite{hen11}, 
which is instrumental for understanding the nuclear symmetry energy~\cite{roc13}. 
Furthermore, not only light-ion but heavy-ion charge-exchange reactions have become an effective probe for investigating 
the multipole excitations, which the nucleonic probes are difficult to study~\cite{len19}. 
Despite the experimental advances, 
most of the theoretical studies have been mostly restricted to spherical nuclei 
except some attempts~\cite{yos10,yos13b,sca14,kor15} though $\Delta T_z=0$ and $\Delta S=0$.

The nuclear shape deformation brings about a characteristic feature in the GRs; 
peak splitting of the GDR, which is caused by the different frequencies of oscillation 
along the long and short axes, has been observed in experiments~\cite{ber75}. 
The splitting of the distribution of the strengths has also been investigated in the isoscalar (IS) giant multipole resonances 
represented as $\Delta T=0$, which is another branch of the GRs~\cite{har01}. 
For the monopole $\Delta L=0$ resonance, the spitting is due to the coupling to the 
$\Delta L_z=0$ component of the quadrupole $\Delta L=2$ resonance~\cite{ume18}, 
which manifests the breaking of the rotational symmetry in the intrinsic frame. 

The present work aims to provide a consistent and systematic description 
of all three modes $\Delta T_z=0,\pm1$ of IV excitations for both electric $\Delta S=0$ and magnetic $\Delta S=1$ types 
in a single framework, and to study the spitting of the distribution of the strengths 
according to $\Delta T_z$ and $\Delta L_z$ or $\Delta J_z$ associated with deformation in isospin space and real space. 
Thus, I consider open-shell nuclei where the nuclear deformation occurs in the ground state 
after demonstrating that the present framework describes the IV responses in spherical nuclei. 
I use a nuclear energy-density-functional (EDF) method: 
a theoretical model being capable of handling nuclides with arbitrary mass numbers~\cite{ben03,nak16},

This paper is organized in the following way: 
the theoretical framework for describing the nuclear responses is given in Sec.~\ref{model} and 
the detail of the numerical procedures is also given; 
Sec.~\ref{result} is devoted to the numerical results and discussion based on the model calculation; 
non-spin flip electric-type excitations and spin-flip magnetic-type excitations are discussed in Sec.~\ref{ele} and Sec.~\ref{mag}, 
respectively; then, a summary is given in Sec.~\ref{summary}.

\section{Theoretical model}\label{model}

\subsection{KSB and QRPA calculations}

Since the details of the formalism can be found in Refs.~\cite{yos08,yos13b,yos13,yos13e}, 
here I briefly recapitulate the basic equations relevant to the present study. 
In the framework of the nuclear EDF method I employ, 
the ground state of a mother (target) nucleus is described by solving the 
Kohn--Sham--Bogoliubov (KSB) equation~\cite{dob84}:
\begin{align}
\sum_{s^\prime}
\begin{bmatrix}
h^q_{s s^\prime}(\br)-\lambda^{q}\delta_{s s^\prime} & \tilde{h}^q_{s s^\prime}(\br) \\
\tilde{h}^q_{s s^\prime}(\br) & -h^q_{s s^\prime}(\br)+\lambda^q\delta_{s s^\prime}
\end{bmatrix}
\begin{bmatrix}
\varphi^{q}_{1,\alpha}(\br s^\prime) \\
\varphi^{q}_{2,\alpha}(\br s^\prime)
\end{bmatrix} \notag \\
= E_{\alpha}
\begin{bmatrix}
\varphi^{q}_{1,\alpha}(\br s) \\
\varphi^{q}_{2,\alpha}(\br s)
\end{bmatrix}, \label{HFB_eq}
\end{align}
where 
the single-particle and pair Hamiltonians, $h^q_{s s^\prime}(\br)$ and $\tilde{h}^q_{s s^\prime}(\br)$, 
are given by the functional derivative of the EDF with respect to the particle density and the pair density, respectively. 
An explicit expression of the Hamiltonians is found in the Appendix of Ref.~\cite{kas21}. 
The superscript $q$ denotes 
$\nu$ (neutron, $ t_z= 1/2$) or $\pi$ (proton, $t_z =-1/2$). 
The average particle number is fixed at the desired value by adjusting the chemical potential $\lambda^q$. 
Assuming the system is axially symmetric, 
the KSB equation (\ref{HFB_eq}) is block diagonalized 
according to the quantum number $\Omega$, the $z$-component of the angular momentum. 

The excited states $| i \rangle$ are described as 
one-phonon excitations built on the ground state $|0\rangle$ of the mother nucleus as 
\begin{align}
| i \rangle &= \hat{\Gamma}^\dagger_i |0 \rangle, \\
\hat{\Gamma}^\dagger_i &= \sum_{\alpha \beta}\left\{
X_{\alpha \beta}^i \hat{a}^\dagger_{\alpha}\hat{a}^\dagger_{\beta}
-Y_{\alpha \beta}^i \hat{a}_{\bar{\beta}}\hat{a}_{\bar{\alpha}}\right\},
\end{align}
where $\hat{a}^\dagger$ and $\hat{a}$ are 
the quasiparticle (qp) creation and annihilation operators that 
are defined in terms of the solutions of the KSB equation (\ref{HFB_eq}) with the Bogoliubov transformation.
The phonon states, the amplitudes $X^i, Y^i$ and the vibrational frequency $\omega_i$, 
are obtained in the quasiparticle-random-phase approximation (QRPA): the linearized time-dependent density-functional theory for superfluid systems~\cite{nak16}. 
The EDF gives the residual interactions entering into the QRPA equation. 
For the axially symmetric nuclei, the QRPA equation 
is block diagonalized according to the quantum number $K=\Omega_\alpha + \Omega_\beta$. 

\subsection{Numerical procedures}

I solve the KSB equation in the coordinate space using cylindrical coordinates
$\boldsymbol{r}=(\varrho,z,\phi)$.
Since I assume further the reflection symmetry, only the region of $z\geq 0$ is considered. 
I use a two-dimensional lattice mesh with 
$\varrho_i=(i-1/2)h$, $z_j=(j-1)h$ ($i,j=1,2,\dots$) 
with a mesh size of
$h=0.6$ fm and 25 points for each direction. 
The qp states are truncated according to the qp 
energy cutoff at 60 MeV, and 
the qp states up to the magnetic quantum number $\Omega=23/2$
with positive and negative parities are included. 
I introduce the truncation for the two-quasiparticle (2qp) configurations in the QRPA calculations,
in terms of the 2qp-energy as 70 MeV.

For the normal (particle--hole) part of the EDF,
I employ the SkM* functional~\cite{bar82}. 
For the pairing energy, I adopt the so-called mixed-type interaction:
\be
V_{\rm{pair}}^{q}(\br,\br^\prime)=V_0
\left[ 1-\frac{\rho(\br)}{2\rho_0} \right]
\delta(\br-\br^\prime)
\ee
with $\rho_0=0.16$ fm$^{-3}$, and $\rho(\br)$ being the isoscalar (matter) particle density. 
I use the parameter $V_0$ as fixed in the previous studies: 
$V_0=-275$ MeV fm$^3$ for the Mg and Si isotopes~\cite{pea16}, 
$V_0=-240$ MeV fm$^3$ for the Ni, Zr, and Pb isotopes~\cite{yos10}.
For the pairing energy of the Sm isotopes, I adopt the one in Ref.~\cite{yam09}
that depends on both the IS and IV densities, 
in addition to the pair density, with the parameters given in Table~III of Ref.~\cite{yam09}. 
The same pair interaction is employed for the dynamical pairing in the QRPA calculation 
and for the $S=0$ and $S=1$ proton--neutron-pairing in the pnQRPA calculation, 
while the linear term in the IV density is dropped. 
Note that the pnQRPA calculations including the dynamic spin-triplet pairing with 
more or less the same strength as the spin-singlet pairing 
describe well the characteristic low-lying Gamow--Teller 
strength distributions in the light $N \simeq Z$ nuclei~\cite{fuj14,fuj15,fuj19}, and the $\beta$-decay half-lives of neutron-rich Ni isotopes~\cite{yos19}.
Furthermore, 
the present theoretical framework describes well the measured giant resonances in light, medium-heavy, and heavy nuclei~\cite{yos13b,nak11,yos11b,yos11c,gup15,gup15e,gup16,pea16,yos21}, 
and low-lying collective modes of vibration~\cite{yos09a,yos13b,yos16,wat16,zha19,yos21b}.

\section{results and discussion}\label{result}

\subsection{Electric modes: Non-spin-flip excitations}\label{ele}

I consider the response to the IV operators defined by
\begin{align}
	\hat{F}_{LK\mu}^{({\rm e})}
	=& \dfrac{1}{\sqrt{2}}\sum_{s s^\prime}\sum_{tt^\prime}\int \dd \br f(r)Y_{LK}(\hat{r})\delta_{s s^\prime}\langle t^\prime|\tau_\mu|t\rangle \notag \\
	& \times \hat{\psi}^\dagger(\br s^\prime t^\prime)\hat{\psi}(\br st), \label{ele_op}
\end{align}  
where $\hat{\psi}^\dagger(\br st), \hat{\psi}(\br st)$ represent the nucleon field operators, 
and $\vec{\tau}=(\tau_{+1},\tau_0,\tau_{-1})$ denotes the spherical components of the Pauli matrix of isospin. 
I take $f(r)=\sqrt{4\pi}$ for the Fermi (F, $L=0$) transition, while $r^2$ for the monopole (M, $L=0$) and quadrupole (Q, $L=2$) transitions.

\subsubsection{Spherical nuclei}\label{ele_sph}

\begin{figure*}[t]
\includegraphics[scale=0.3]{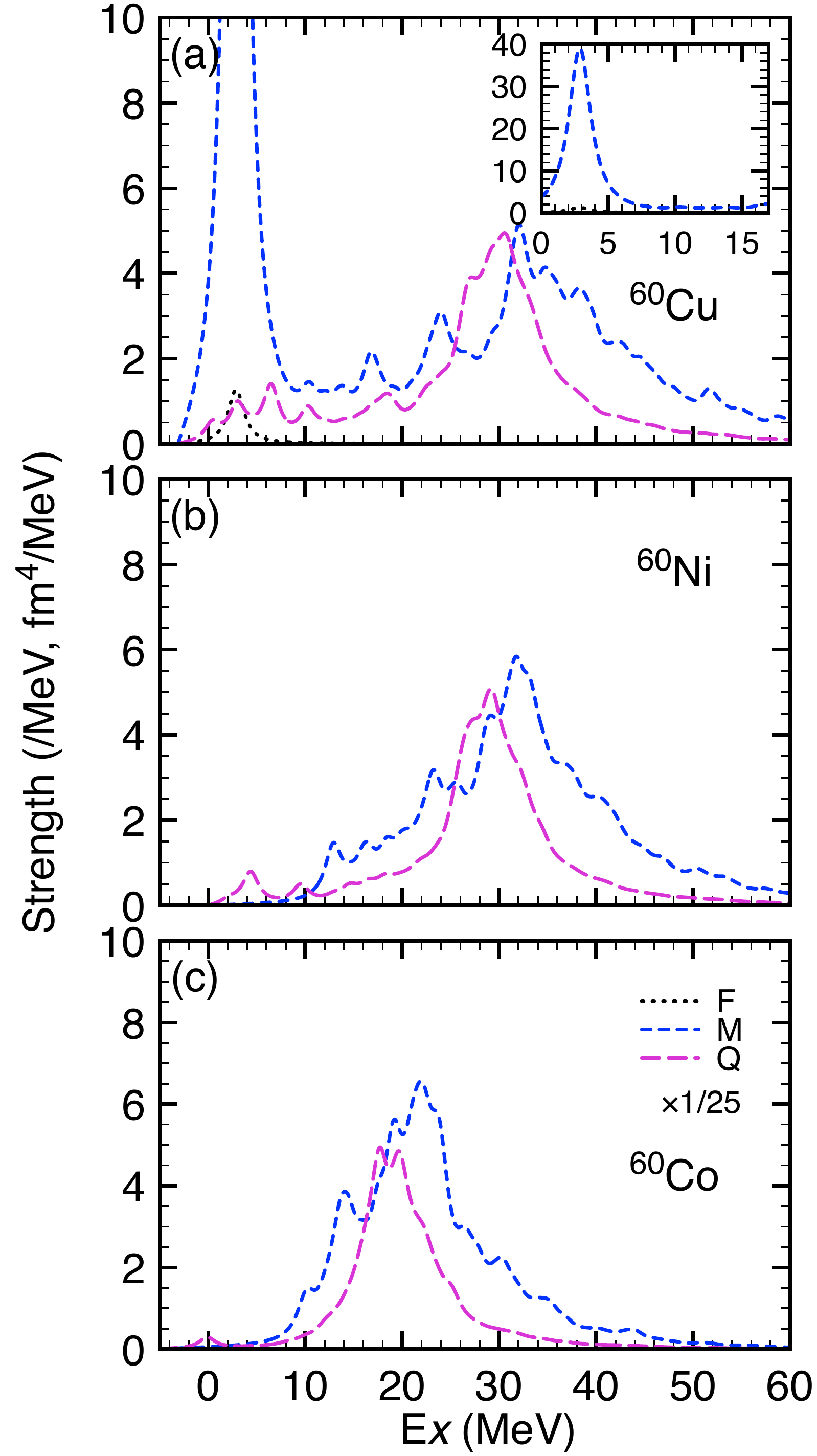}
\includegraphics[scale=0.3]{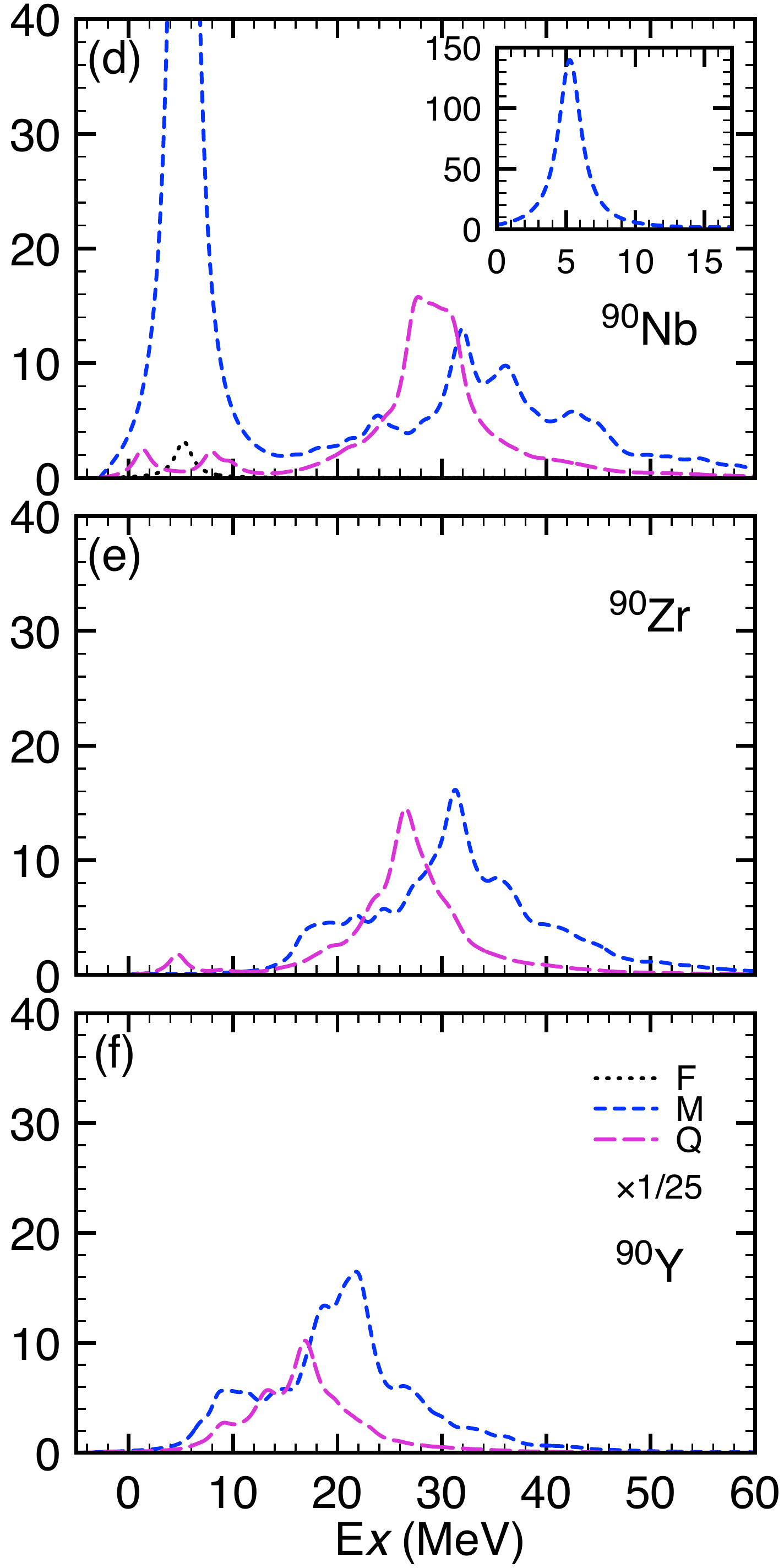}
\includegraphics[scale=0.3]{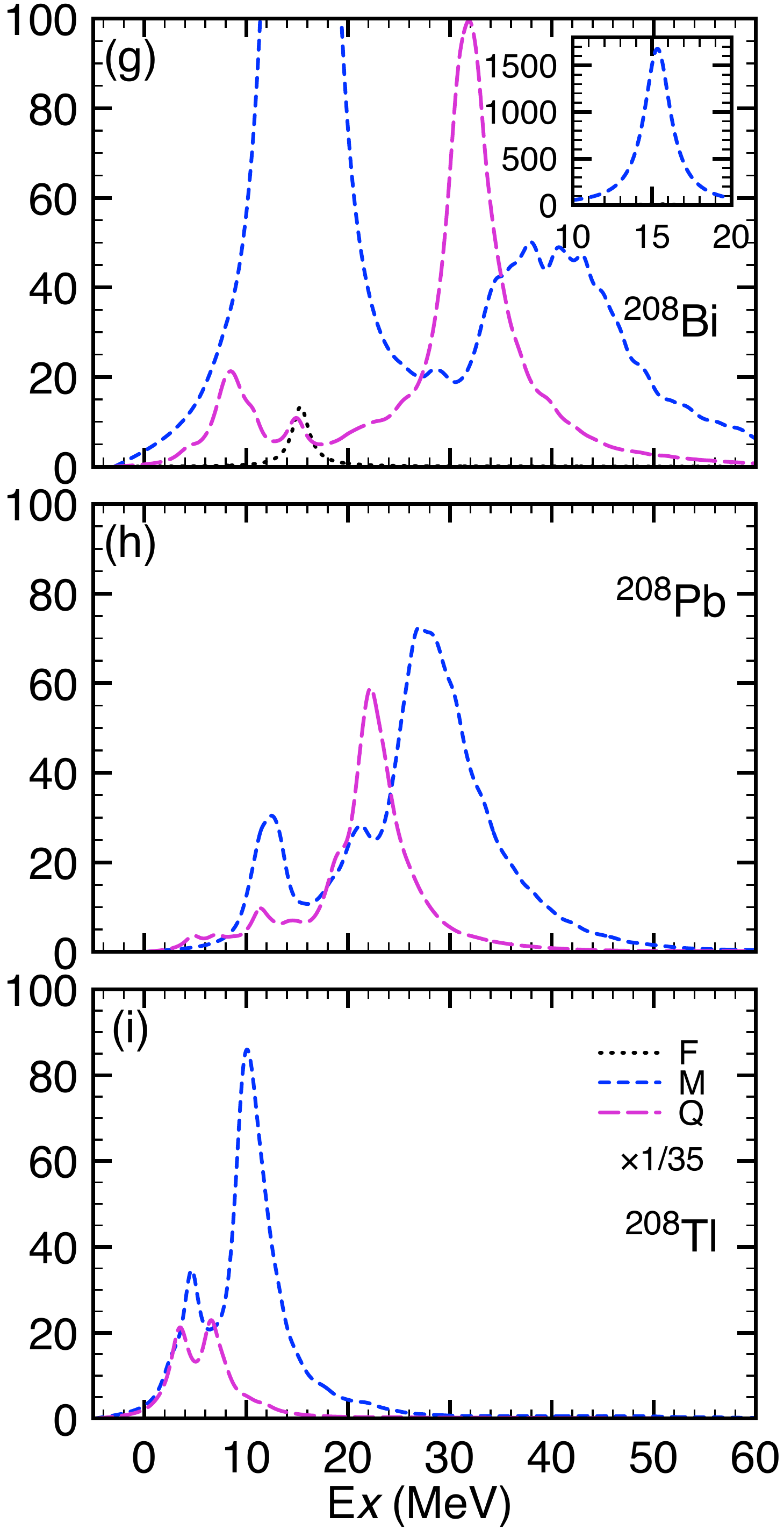}
\caption{\label{fig:ele_sph}
Transition strengths of the non-spin-flip excitations in the 
$\mu=-1$ [(a), (d), (g)], $\mu=0$ [(b), (e), (h)], and $\mu=+1$ channels [(c), (f), (i)]. 
The Fermi (F), monopole (M), and quadrupole (Q) strengths are shown by the dotted, dashed, and long-dashed lines, 
respectively. The quadrupole strengths are multiplied by $1/25$ ($1/35$ for $^{208}$Bi, Pb, Tl). 
The excitation energies $E_x$ are with respect to the ground state of the daughter nucleus. 
 }
\end{figure*}

Before investigating deformed nuclei, I study the IV giant resonances in some spherical nuclei, 
where the experimental data are available. 
Figure~\ref{fig:ele_sph} shows the transition-strength distributions in $^{60}$Ni, $^{90}$Zr, and $^{208}$Pb as an example of spherical nuclei:  
\begin{align}
S_L^\mu(E)&=\sum_K \dfrac{\dd B(E,F_{LK\mu})}{\dd E}, \label{eq:strength1}\\
\dfrac{\dd B(E,F_{LK\mu})}{\dd E} &=\dfrac{2E\gamma}{\pi}\sum_{i}
\dfrac{\tilde{E}_{i} |\langle i|\hat{F}^{({\rm e})}_{LK\mu}|0 \rangle|^{2}}{(E^{2}-\tilde{E}_{i}^{2})^{2}+E^{2}\gamma^{2}}, \label{eq:strength}
\end{align}
where $\tilde{E}_{i}^{2}=(\hbar \omega_{i})^{2}+\gamma^{2}/4$~\cite{BM2}. 
The smearing width $\gamma$ is set to 2 MeV, which is supposed to simulate
the spreading effect, $\Gamma^\downarrow$, missing in the QRPA. 
For the charge-exchange modes of excitation, the excitation energy with respect to the ground state 
of the mother nucleus is evaluated by replacing $E$ by $E\pm (\lambda^\nu-\lambda^\pi)$ for the $\mu=\pm 1$ channel~\cite{yuk20}. 
Furthermore, in plotting the strength distributions with respect to the ground state of the daughter nucleus, 
the mass difference between the mother and daughter is considered by using AME2020~\cite{Huang_2021,Wang_2021}: 
the ground-state $Q$ value is $-6.1$ MeV and $-2.8$ MeV in $^{60}{\rm Cu}$ and ${}^{60}{\rm Co}$ with respect to ${}^{60}{\rm Ni}$, 
$-6.1$ MeV and $-2.2$ MeV in ${}^{90}{\rm Nb}$ and ${}^{90}{\rm Y}$ with respect to ${}^{90}{\rm Zr}$, and 
$-2.9$ MeV and $-5.0$ MeV in ${}^{208}{\rm Bi}$ and ${}^{208}{\rm Tl}$ with respect to ${}^{208}{\rm Pb}$.

A striking feature one sees in the $\mu=-1$ channel is 
the concentration of the monopole strength in the isobaric analog resonance (IAR). 
I find 53\%, 68\%, and 85\% of the total strength in the IAR in $^{60}$Cu, $^{90}$Nb, and $^{208}$Bi, respectively, as summarized in Tab.~\ref{tab:E_sph}. 
It is noted that the summed strengths excluding the IAR are given in the parenthesis in Tab.~\ref{tab:E_sph}. 
A similar trait was also found in the early investigation~\cite{aue83}.
In the high-frequency region, the peak energy of the monopole resonance is higher than the quadrupole resonance. 
This is also the case in the $\mu=0$ and $\mu=+1$ channels.
Furthermore, 
the strengths are spread out over a wider energy region for the monopole resonance; 
the width of the IVGMR is larger than that of the IVGQR. 

Table~\ref{tab:E_sph} lists the summed strengths for the monopole and quadrupole excitations. 
One can see that the present calculation satisfes the model-independent non-energy weighted sum rule for the charge-exchange modes~\cite{BM2}: 
\begin{align}
m_L^{-1}&-m_L^{+1} 
\notag\\
&=\left\{
\begin{array}{ll}
N-Z & {\rm F} \\
\dfrac{2L+1}{4\pi}(N\langle r^4\rangle_\nu - Z\langle r^4\rangle_\pi) & {\rm M,Q}
\end{array}
\right.,
\label{eq:NEWSR}
\end{align}
where $\langle \cdots \rangle_{\nu (\pi)}$ stands for the expectation value evaluated for neutrons (protons) in the ground state of the mother nucleus, and 
\begin{align}
m_L^{\mu} = \int \dd E S_L^{\mu}(E).
\end{align}
In these nuclei, 
$m^{-1}$ is always larger than $m^{+1}$
because $\langle r^4\rangle$ for neutrons is slightly larger than that for protons. 
The monopole and quadrupole excitations are primarily built of a coherent particle--hole configurations of $2\hbar \omega_0$ excitation, 
and the high-frequency resonance is such a mode of excitation. 
However, the $0 \hbar \omega_0$ excitation can also be involved. 

For the monopole excitations, the $\nu 2p_{3/2} \to \pi 2p_{3/2}$ and $\nu 1g_{9/2} \to \pi 1g_{9/2}$ excitation generates the IAR of $^{60}$Ni and $^{90}$Zr, 
while the $0 \hbar \omega_0$ excitation is strongly supressed in the $\mu=0$ and $+1$ channels due to the Pauli blocking. 
Therefore, the summed strength $m^{-1}$ excluding the IAR has a similar value to $m^0$ and $m^{+1}$, 
which indicates that the higher-energy monopole strengths in the $\mu=-1$ channel represent the $2\hbar \omega_0$ excitation.
In $^{208}$Pb, no $0\hbar \omega_0$ excitation is available in the $\mu=0$ and $\mu=+1$ channels as in $^{60}$Ni and $^{90}$Zr, 
and the summed strength $m^{-1}$ excluding the IAR has a similar value to $m^0$. 
However, 
the number of particle--hole configurations in the $\mu=+1$ channel and the $m^{+1}$ value are smaller 
since the Fermi levels of neutrons and protons are located apart by $N=1$. 

\begin{table}[t]
\caption{\label{tab:E_sph} 
Summed monopole and quadrupole strengths, and comparison with the non-energy-weighted sum rule (NEWSR) values, given in units of fm$^4$. 
$\langle r^4\rangle$ for neutrons (protons) is 276 (257) fm$^4$, 444 (421) fm$^4$, and 1284 (1114) fm$^4$ in $^{60}$Ni, $^{90}$Zr, and $^{208}$Pb, 
respectively.
The summed monopole strengths excluding the strength of the IAR are given in the parenthesis. 
}
\begin{ruledtabular}
\begin{tabular}{lccccc}
& $m_L^{-1}$ & $m_L^{+1}$ & $m_L^0$ & $m_L^{-1}-m_L^{+1}$ & NEWSR \\
 \hline
$^{60}$Ni & & & & & \\
 $L=0$ & 228.3 (107.1) & 97.00 & 107.3 & 131.3 & 131.1 \\
 $L=2$ & 1901 & 1242 & 1546 & 658.6 & 655.4 \\
$^{90}$Zr & & & & & \\
 $L=0$ & 639.2 (202.2) & 212.2 & 217.8 & 427.0 & 425.8 \\
 $L=2$ & 4427 & 2289 & 3259 & 2138 & 2129 \\
$^{208}$Pb & & & & & \\
 $L=0$ & 6210 (952.3) & 599.6 & 1044 & 5610 & 5607 \\
 $L=2$ & 33734 & 5681 & 16063 & 28053 & 28036
\end{tabular}
\end{ruledtabular}
\end{table}

The quadrupole excitation is more involved. 
In $^{60}$Ni, the $1f_{7/2}\to 1f_{5/2}$ excitation is available in all the channels. 
The $\nu 1p_{3/2}\to \pi 1p_{3/2}$ excitation participates in the low-lying $2^+$ excitation in the $\mu=-1$ channel, 
and the $1p_{3/2}\to 1p_{1/2}$ excitation further contribute to generate the $2^+$ excitation in the $\mu=-1$ and $\mu=0$ channels. 
Thus, the $2^+$ states appear in low energy with the transition strengths dependent on $\mu$. 
In $^{90}$Zr, the $1g_{9/2} \to 1g_{7/2}$ excitation generates the low-lying $2^+$ excitation in the $\mu=0$ and $\mu=-1$ channel. 
Furthermore, the $\nu 1g_{9/2} \to \pi 1g_{9/2}$ excitation participates in the low-lying $2^+$ excitation in the $\mu=-1$ channel. 
Therefore, one sees the strengths in low energy, while there are no strengths in the $\mu=+1$ channel 
since the $0 \hbar \omega_0$ excitation is not available. 
In $^{208}$Pb, both $0\hbar \omega_0$ and $2\hbar \omega_0$ excitations generate the $2^+$ excitation in the $\mu=-1$ channel, 
acquiring a large strength. 
In the $\mu=0$ channel, the $\pi 1h_{11/2} \to \pi 1h_{9/2}$ and $\nu 1i_{13/2}\to \nu 1i_{11/2}$ excitations as well as 
the $2\hbar \omega_0$ excitation generate the $2^+$ excitation. 
However, the $0 \hbar \omega_0$ excitation is unavailable in the $\mu=+1$ channel. 
Therefore, the transition strengths in the $\mu=+1$ channel are smaller than in the other channels as in the monopole case. 

Here, I compare the calculated strength distributions with the available experimental data. 
A systematic study of the charge-exchange $(\pi^{\pm},\pi^0)$ reaction reveals the IVGMR in medium-mass and heavy nuclei~\cite{ere86}: 
the excitation energy of the IVGMR measured using the $^{208}{\rm Pb}(\pi^+,\pi^0){}^{208}{\rm Bi}$ reaction is $37.2 \pm 3.5$ MeV, 
while $E_x=12.0 \pm 2.8$ MeV in $^{208}{\rm Pb}(\pi^-,\pi^0){}^{208}{\rm Tl}$. 
The excitation energy in lighter nuclei is $E_x=35.6 \pm 2.8$ and $25.2 \pm 1.7$ MeV for $^{60}{\rm Cu}$ and $^{60}{\rm Co}$, 
and $E_x=34.6 \pm 2.9$ and $22.0 \pm 2.0$ MeV for $^{90}{\rm Nb}$ and $^{90}{\rm Y}$. 
The inelastic electron scattering experiment suggests the resonance around 33 MeV in $^{208}$Pb as the IVGMR ~\cite{pit74}, 
though this is $\sim$5 MeV higher than the average of $E_{T-1}$ and $E_{T+1}$ obtained using the charge-exchange reaction.
The nuclear reactions have also been employed to measure the IVGMR. 
The ${\rm Pb}({}^{3}{\rm He},tp){\rm Bi}$ reaction indicates the location of the IVGMR or spin monopole resonance at $30\textendash45$ MeV 
with respect to the ground state of Pb~\cite{zeg00}. 
The IVGMR measured using the $({}^{7}{\rm Li},{}^{7}{\rm Be})$ reaction is found at $20 \pm 2$ MeV in $^{60}{\rm Co}$~\cite{nak99}. 
In most cases, the present calculation describes well the location of the IVGMR. 

The IVGQR has been found around $130 \times A^{-1/3}$ MeV in the $\mu=0$ channel~\cite{har01}. 
In $^{208}$Pb, the excitation energy is 20--23 MeV~\cite{lei81,sch88,dal92,hen11}. 
The IVGQR in $^{90}$Zr is located around 26--27 MeV~\cite{fuk76,god94}. 
The present calculation employing the SkM* functional reproduces well these experimental data. 
The $({}^{13}{\rm C},{}^{13}{\rm N})$ reaction has been employed to locate the IVGQR in $^{60}{\rm Co}$, 
and it is found at $E_x=20 \pm 2$ MeV~\cite{ich02}. 
The calculation is in remarkable agreement with the experiment, as shown in Fig.~\ref{fig:ele_sph}(c)

\subsubsection{Deformation effects}\label{ele_def}

\begin{figure}[t]
\begin{center}
\includegraphics[scale=0.26]{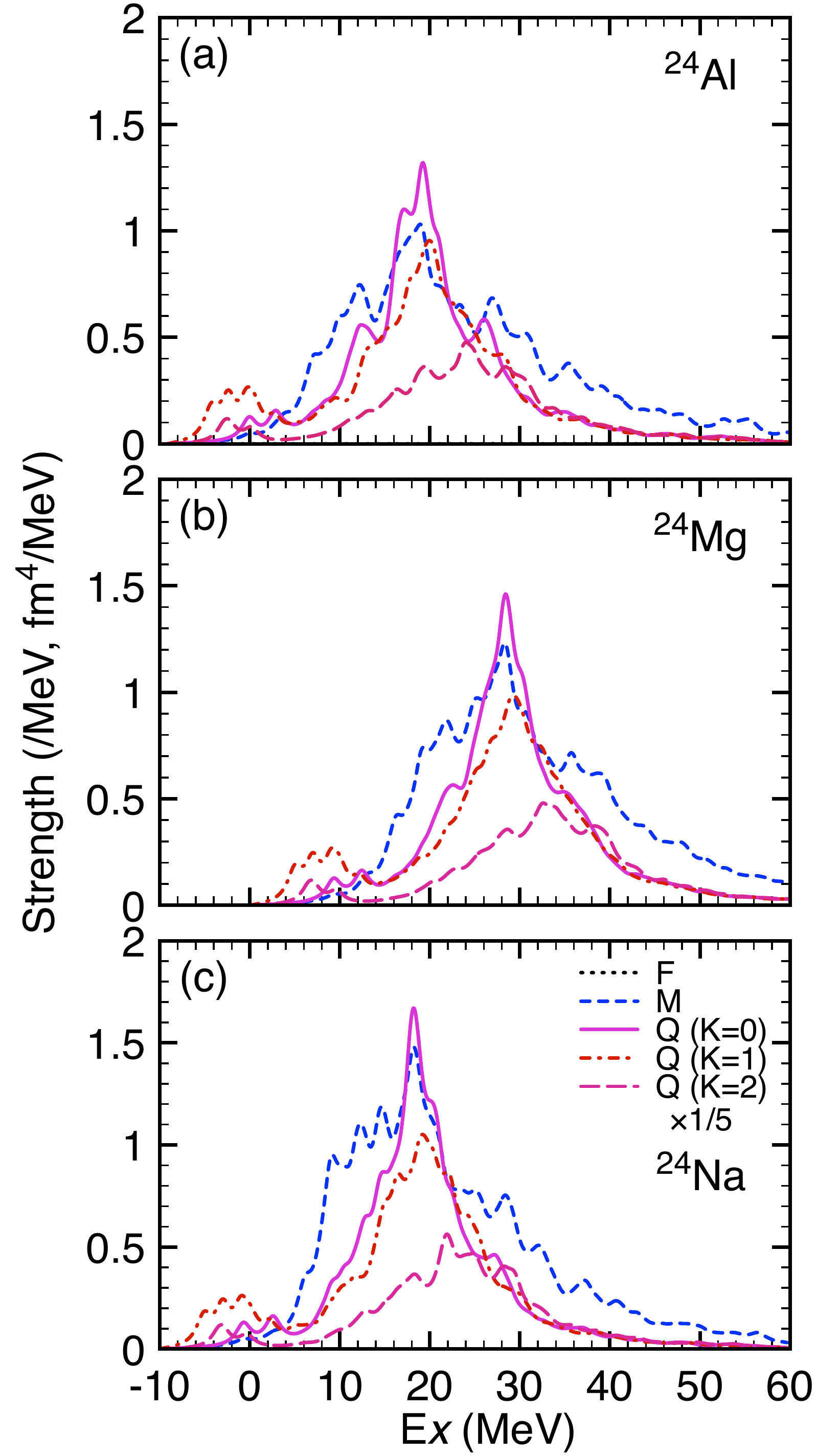}
\includegraphics[scale=0.26]{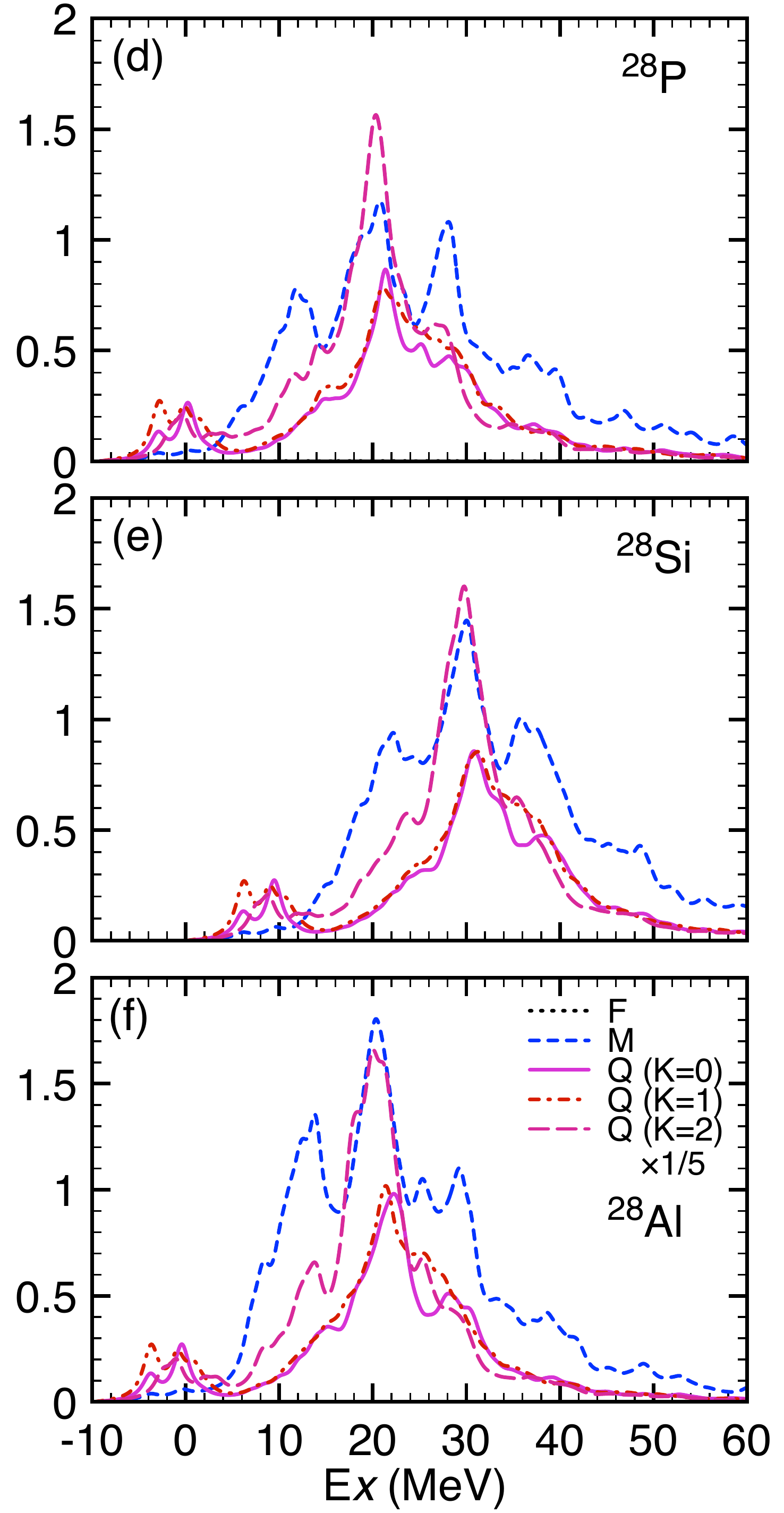}
\caption{\label{fig:ele_def} 
As Fig.~\ref{fig:ele_sph} but for the deformed $^{24}$Mg and $^{28}$Si nuclei. 
Instead of showing the total strengths, those for each $K$ component are shown for the quadrupole excitations. 
The quadrupole strengths are multiplied by $1/5$. 
}
\end{center}
\end{figure}

I am going to investigate the deformation effects. 
Figure \ref{fig:ele_def} shows the transition-strength distributions in $^{24}$Mg and $^{28}$Si as an example of light deformed nuclei. 
As discussed in Refs.~\cite{gup15,gup15e,gup16,pea16}, the ground state is prolately deformed and oblately deformed 
with the deformation parameter $\beta_2=0.39$ and $-0.22$ in $^{24}$Mg and $^{28}$Si, respectively. 
Since these nuclei have the same number of protons and neutrons, the Fermi transition strength is weak. 
A characteristic feature of these $N=Z$ nuclei is that the transition strength distributions in the three channels are similar to each other.
For the dipole case, this characteristic trait has been discussed in Ref.~\cite{yos20}. 
Without the Coulomb potential, one cannot distinguish the motion of protons and neutrons in $N=Z$ nuclei, 
and the isotripet states are degenerated. 
However, the Coulomb potential slightly expands the proton distribution, which leads to the asymmetry, 
as expected by the sum rule~(\ref{eq:NEWSR}). 
A simple RPA analysis for a single normal mode employing the separable interaction 
gives the relation for the summed transition strengths as~\cite{BM2} 
\begin{equation}
\dfrac{1}{2}(m^{-1}+m^{+1})=\left[1+O\left(\frac{N-Z}{A}\right)\right]m^0.
\label{eq:strength_rel}
\end{equation}

In deformed nuclei, 
the $K$-splitting occurs for the multipole modes of excitation, and thus 
the sum rule (\ref{eq:NEWSR}) for the quadrupole excitation 
is generalized by replacing $\langle r^4\rangle$ with 
\begin{align}
\begin{array}{ll}
\dfrac{5}{4}\langle 4z^4 + \rho^4 - 4\rho^2z^2\rangle & {\rm Q} (K=0) \\
\dfrac{15}{2}\langle \rho^2z^2\rangle & {\rm Q} (K=\pm 1) \\
\dfrac{15}{8}\langle \rho^4\rangle & {\rm Q} (K=\pm 2)
\end{array}
\label{eq:NEWSR_def}
\end{align}
depending on the $K$ quantum number. 
Table~\ref{tab:E_def} summarizes the summed strengths in $^{24}$Mg and $^{28}$Si, 
and the NEWSR values taking the nuclear deformation into account (\ref{eq:NEWSR_def}). 
One finds that in both nuclei the relation (\ref{eq:strength_rel}) holds accurately. 
It should be noted that the relation (\ref{eq:strength_rel}) is model dependent. 
However, the present selfconsistent model satisfies the simple relation, suggesting a rather generous rule for 
the IV excitations. 

\begin{table}[t]
\caption{\label{tab:E_def} 
As Tab.~\ref{tab:E_sph} for $^{24}$Mg and $^{28}$Si.
}
\begin{ruledtabular}
\begin{tabular}{lccccc}
& $m_L^{-1}$ & $m_L^{+1}$ & $m_L^0$ & $m_L^{-1}-m_L^{+1}$ & NEWSR \\
 \hline
$^{24}$Mg & & & & & \\
 $L=0$         & 23.27 & 29.25 & 26.20 & $-5.98$ & $-6.01$ \\
 $L=2, K=0$ & 86.94 & 95.77 & 91.40 & $-8.83$ & $-8.93$ \\
 $L=2, K=1$ & 80.47 & 87.56 & 84.32 & $-7.09$ & $-7.18$ \\
 $L=2, K=2$ & 45.32 & 48.68 & 47.04 & $-3.36$ & $-3.39$ \\
$^{28}$Si & & & & & \\
 $L=0$ & 26.82 & 34.67 & 30.66 & $-7.84$ & $-7.87$ \\
 $L=2, K=0$ & 65.43 & 71.66 & 68.64 & $-6.23$ & $-6.29$ \\
 $L=2, K=1$ & 73.32 & 79.62 & 76.54 & $-6.30$ & $-6.35$ \\
 $L=2, K=2$ & 93.86 & 104.0 & 99.02 & $-10.12$ & $-10.18$
\end{tabular}
\end{ruledtabular}
\end{table}

As mentioned above, the $^{24}$Mg and $^{28}$Si nuclei have different shapes in the ground states: 
prolate deformation in $^{24}$Mg and oblate deformation in $^{28}$Si. 
As a consequence of the prolate (oblate) deformation, distinctive features show up 
in the quadrupole strength distributions in high energy. 
The $K=0$ ($K=2$) states move toward low energy and acquire more considerable strengths in 
the prolately (oblately) deformed configuration. 
Furthermore, the coupling to the $K=0$ component of the IVGQR brings about 
the resonance peak in the IVGMR. 
These features are common to the IS excitation. 
The enhancement of the $K=0$ ($K=2$) strengths in the prolate (oblate) configuration, 
which is also seen in Tab.~\ref{tab:E_def}, 
may be understood by looking at the summed strengths (\ref{eq:NEWSR_def}). 
In a prolately (oblately) deformed state, $\langle z^4 \rangle$ increases (decreases), 
while $\langle \rho^4\rangle$ decreases (increases), though the evaluation of $\langle \rho^2 z^2 \rangle$ 
requires a detail of the density distribution. 

In Ref.~\cite{sco17}, the ${}^{28}{\rm Si}({}^{10}{\rm Be},{}^{10}{\rm B}^*)$ reaction 
has been employed to identify the IVGMR in a deformed nucleus. 
The differential cross-section displays a broad peak ranging from 10 MeV to 30 MeV in $^{28}$Al. 
The present calculation reasonably explains the measurement. 
However, it is not easy to find unique features due to deformation as the strength distribution is spread over a wide energy range.

The coupling between the GMR and the $K=0$ component of the GQR becomes strong in a strongly deformed nucleus, 
which has been investigated for the ISGMR in detail from light to medium-heavy nuclei~\cite{ume18}. 
The deformation effect on the coupling has also been investigated theoretically for the IVGMR~\cite{yos10,yos13b}: 
in the $\mu=0$ channel, the IVGMR shows up at about 30 MeV and the IVGQR around 25 MeV in the Nd and Sm isotopes; 
see  Figs.~6(b) and 6(d) of Ref.~\cite{yos13b}. 
In $^{154}$Sm, which is strongly deformed, a resonance peak appears around 20 MeV and 
one finds clearly the splitting of the monopole strengths~\cite{yos13b}. 
Because the study in Ref.~\cite{yos13b} is restricted to the $\mu=0$ channel, 
I am going to investigate the deformation effect on the coupling in the $\mu=\pm 1$ channels and to see the coupling between 
the GMR and the $K=0$ component of the GQR is a general feature emerging in deformed nuclei.

\begin{figure}[t]
\begin{center}
\includegraphics[scale=0.3]{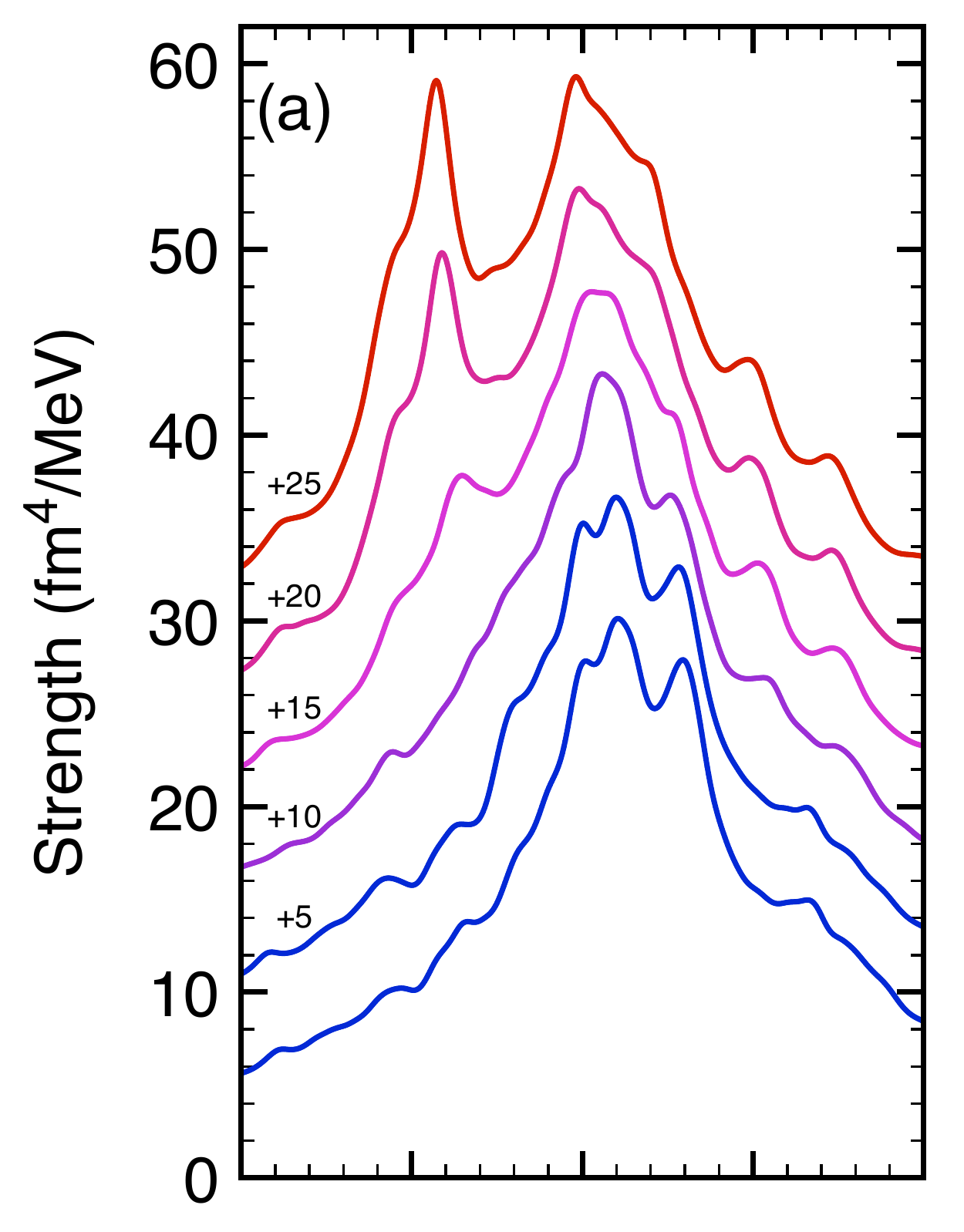}
\includegraphics[scale=0.3]{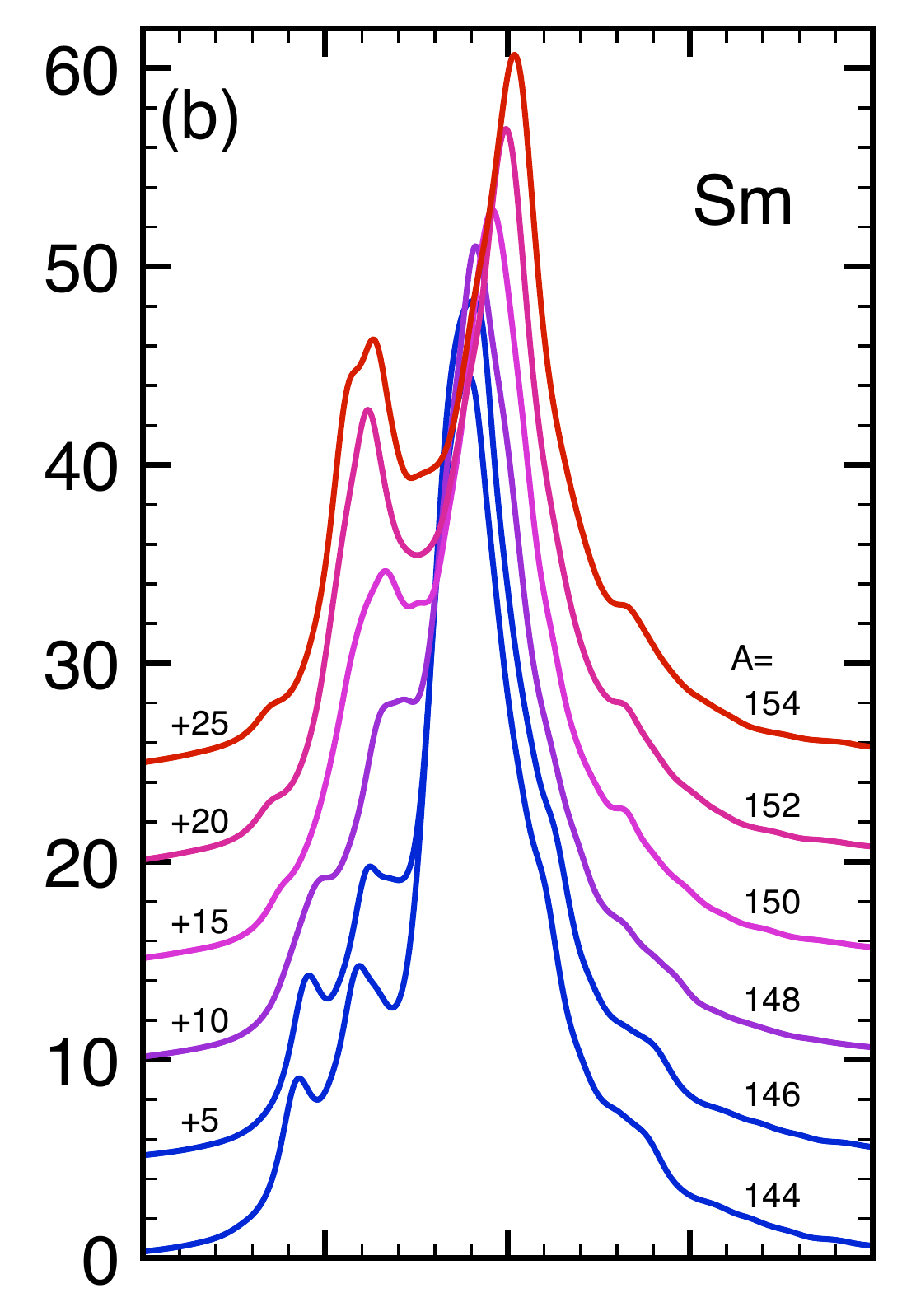} \\
\includegraphics[scale=0.3]{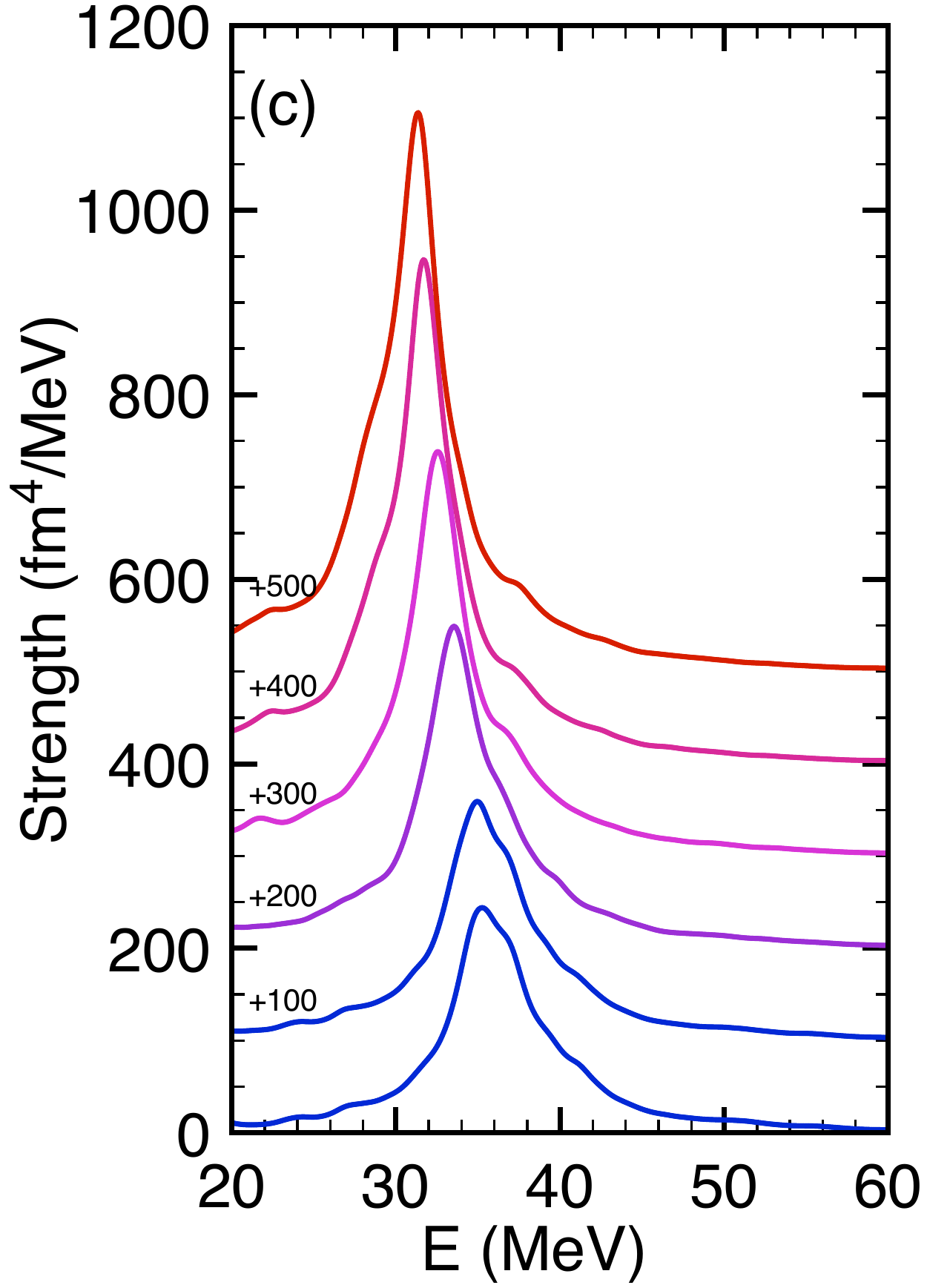}
\includegraphics[scale=0.3]{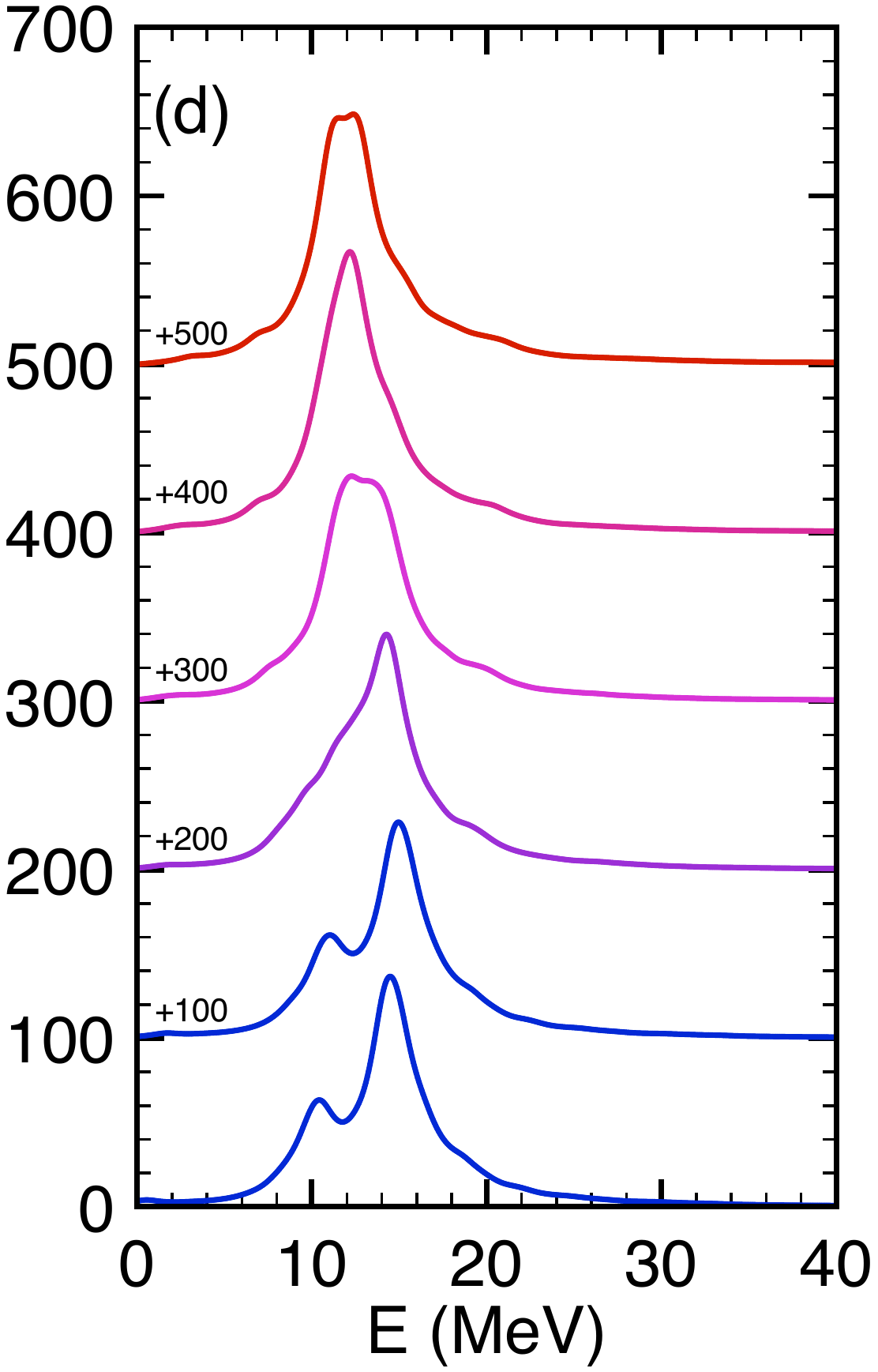}
\caption{\label{fig:Sm_mono} 
Monopole strengths in the (a) $\mu=-1$ channel and (b) $\mu=+1$ channel of the Sm isotopes (shifted), 
and the $K=0$ component of quadrupole strength distributions in the (c) $\mu=-1$ channel and (d) $\mu=+1$ channel (shifted). 
The excitation energies are with respect to the ground state of the target nuclei. 
}
\end{center}
\end{figure}

Figures~\ref{fig:Sm_mono}(a) and \ref{fig:Sm_mono}(b) show the monopole strength distributions in the $\mu=-1$ 
and $\mu=+1$ channels of the Sm isotopes. 
Here, the excitation energies are with respect to the ground state of the targets: the Sm isotopes. 
The IARs are excluded in plotting the strength distribution for the monopole strengths in the $\mu=-1$ channel, 
because most of the strengths are found in the IAR. 
One sees that a lower-energy resonance shows up around 30 MeV in $^{150,152,154}$Sm, 
while there appears a resonance around 40--50 MeV in all the isotopes, which is considered as a primal IVGMR.  
The SkM* functional produces the onset of quadrupole deformation in between $N=84$ and 86, 
and the deformation gradually develops with an increase in the neutron number~\cite{yos11b}. 
The stronger the ground-state deformation, the more enhanced the transition strengths in the lower energy region. 
The $K=0$ component of the quadrupole strengths is shown in Fig.~\ref{fig:Sm_mono}(c). 
One finds that the monopole resonance in low energy is strongly coupled with the $K=0$ component of the IVGQR in the well-deformed isotopes. 

A similar feature can be seen in the $\mu=+1$ channel: 
one sees a resonance around 20-25 MeV in all the Sm isotopes, and a prominent peak appears in $^{152,154}$Sm in low energy at $\sim 10\textendash15$ MeV. 
The $K=0$ component of the IVGQR in these isotopes has a peak around 10--15 MeV, as shown in Fig.~\ref{fig:Sm_mono}(d), 
where the lower-energy resonance of the monopole strengths shows up. 

\subsection{Magnetic modes: Spin-flip excitations}\label{mag}

Here, I consider the response to the IV operators defined by
\begin{align}
	\hat{F}_{J K \mu}^{({\rm m})}
	=& \dfrac{1}{\sqrt{2}}\sum_{ss^\prime}\sum_{tt^\prime}\int \dd \br f(r)[Y_L \otimes \vec{\sigma}]^J_K
	\langle t^\prime|\tau_\mu|t\rangle \notag \\
	& \times \hat{\psi}^\dagger(\br s^\prime t^\prime)\hat{\psi}(\br st), \label{mag_op}
\end{align}  
where $[Y_L \otimes \vec{\sigma}]^J_K = \sum_{\nu \nu^\prime}\langle L\nu 1 \nu^\prime|J K\rangle Y_{L\nu}(\hat{r})\langle s^\prime|\sigma_{\nu^\prime}|s\rangle$ 
with the spherical components of the Pauli spin matrix $\vec{\sigma}=(\sigma_{+1},\sigma_0,\sigma_{-1})$. 
I take $f(r)=\sqrt{4\pi}$ for the GT $(L=0)$ transition, while $r^2$ for the monopole $(L=0)$ and quadrupole $(L=2)$ transitions as in the electric cases. 
The $J=3$ spin-quadrupole (SQ) excitation in the $\mu=0$ channel corresponds to the spin-M3 excitation apart from a factor.

\subsubsection{Spin quadrupole excitations in spherical nuclei}

Since there are plenty of studies on the GT and spin monopole (SM) responses in spherical nuclei, such as in Ref.~\cite{ham00} where 
the Skyme EDF method has been applied to the SM excitations, 
I do not show similar results to Ref.~\cite{ham00}, but rather I focus on the SQ excitations.

\begin{figure}[t]
\begin{center}
\includegraphics[scale=0.26]{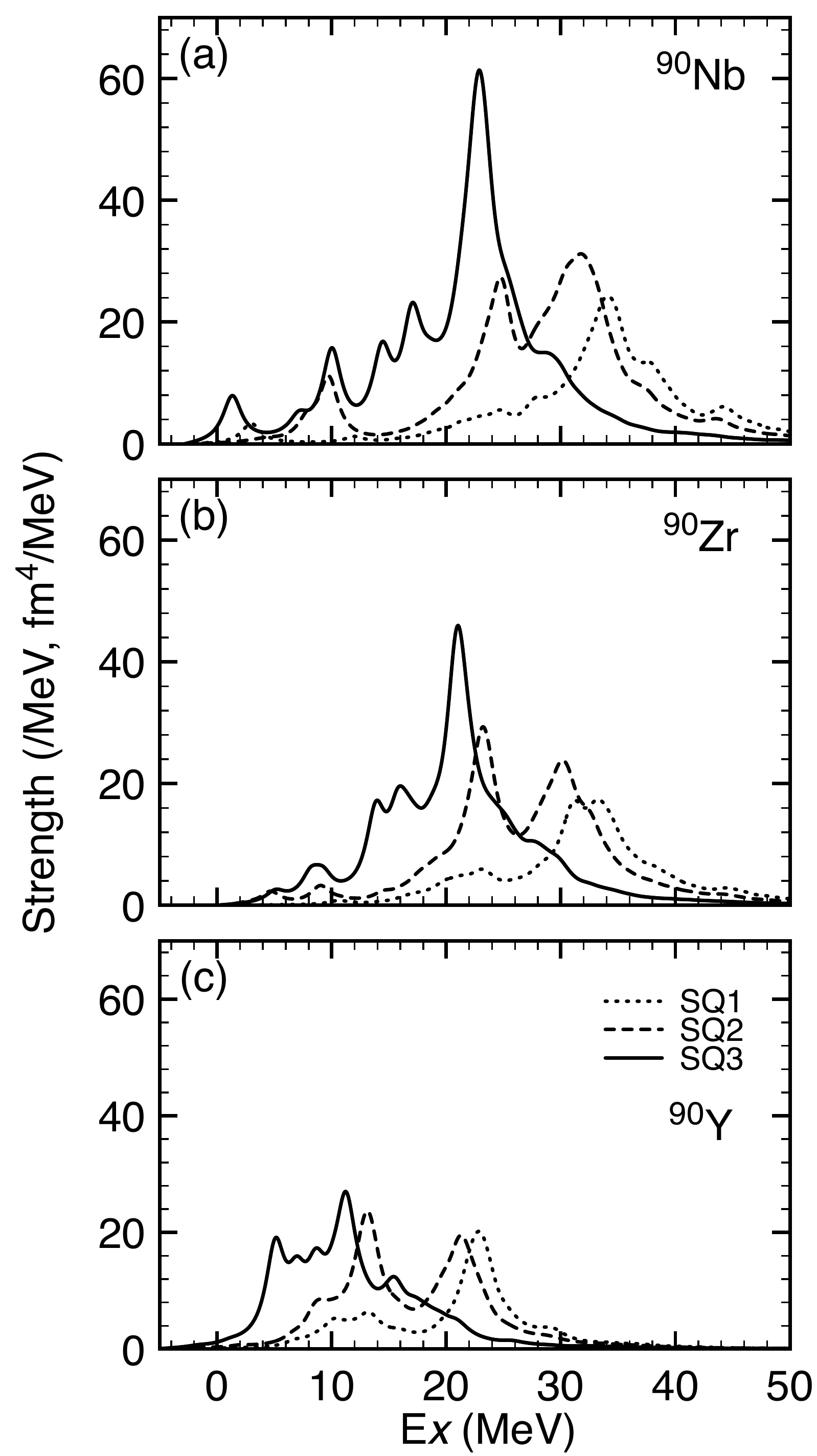}
\includegraphics[scale=0.26]{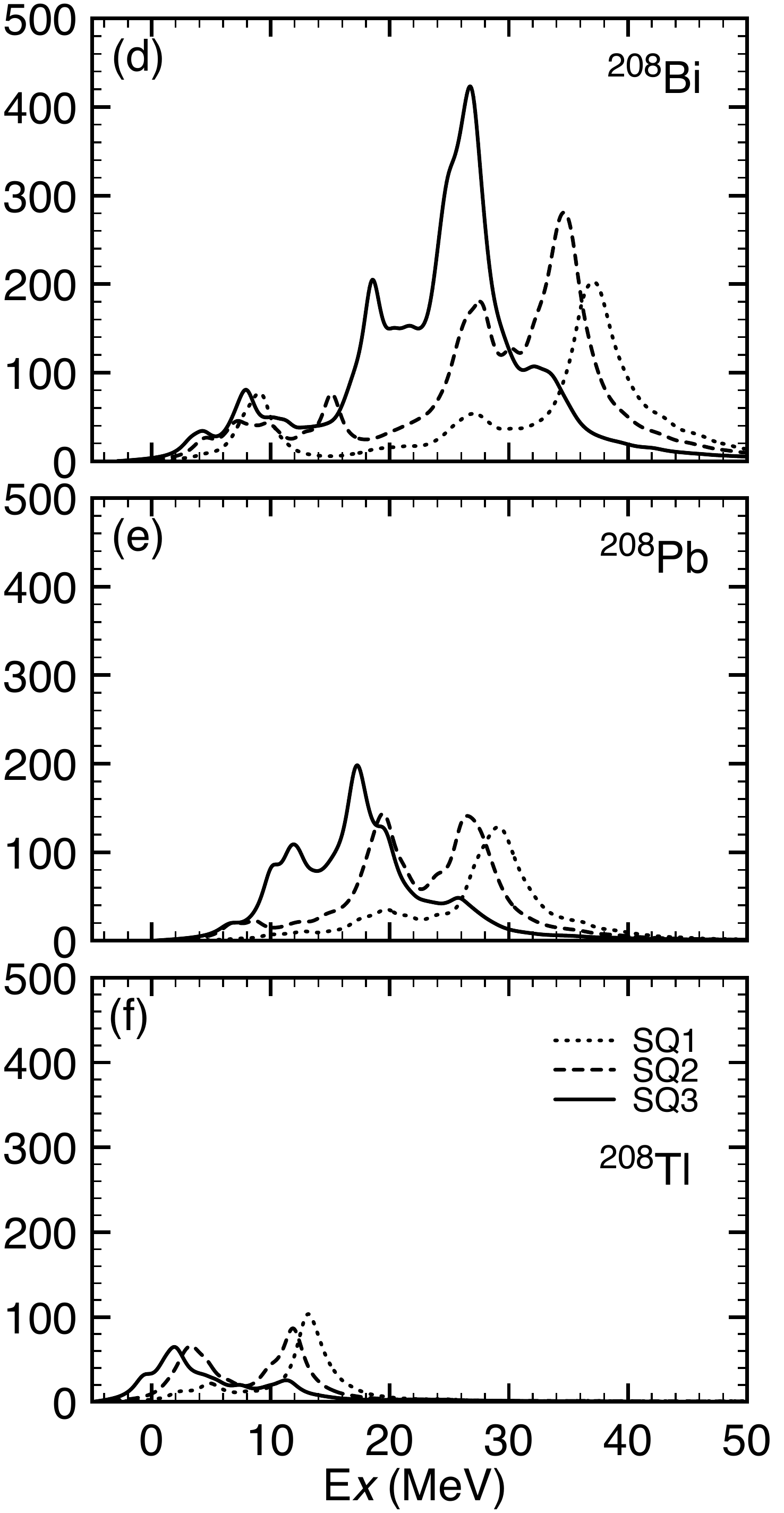}
\caption{\label{fig:sp_qua} 
As Fig.~\ref{fig:ele_sph} but for the spin quadrupole (SQ) strengths. 
The $J=1$, 2, and 3 states are depicted by the dotted, dashed, and solid lines, respectively. 
}
\end{center}
\end{figure}

Figure~\ref{fig:sp_qua} shows the transition-strength distributions in ${}^{90}$Zr 
and ${}^{208}$Pb as an example of spherical nuclei. 
As in the electric cases, the $\mu$-dependence of the strength distribution is more substantial with increasing excess neutrons. 
In $^{90}$Zr, the excitations are mainly built of the $2\hbar \omega_0$ excitation: $N=3 \to 5$ and $N=2 \to 4$. 
Among them, the $1f_{5/2} \to 1h_{11/2}$ excitation with $J=3$ appears in low energy. 
In the $\mu=0$ and $-1$ channels, the $0\hbar \omega_0$ excitation is also possible to occur: 
the particle--hole excitations from the $\nu 1g_{9/2}$ orbital within the $N=4$ shell.  
Furthermore, the $\nu 1g_{9/2} \to \pi 1g_{9/2}$ excitation participates in forming the low-lying states in the $\mu=-1$ channel. 
In $^{208}$Pb, the $\pi 2d_{3/2} \to \nu 2g_{9/2}$ excitation with $J=3$ 
and the $\pi 2d_{3/2} \to \nu 3d_{5/2}$ and $\pi 1h_{11/2} \to \nu1 j_{15/2}$ excitations generate the low energy states in the $\mu=+1$ channel. 
In the $\mu=0$ and $-1$ channels, the $0\hbar \omega_0$ excitation is also available: 
the particle--hole excitations from the $\nu 1i_{13/2}$ orbital in the $N=6$ shell. 
Furthermore, the $\nu 1h_{11/2} \to \pi 1h_{11/2}$ excitation participates in forming the low-lying states in the $\mu=-1$ channel.

In these examples, one sees that the excitation energy of $J=3$ is the lowest and $J=1$ the highest. 
This is already seen in the unperturbed strength distributions and is consistent with the finding in the early study~\cite{aue84}. 
This is partly because the $J=3$ states are constructed by the particle--hole excitation of the orbitals with $(\ell-2)_{j_<}$ and $\ell_{j_>}$, 
whose unperturbed energy is lowered by the spin--orbit interaction. 
This explanation is similar to that quoted for the lowering of the $J=2$ states of the spin dipole excitations~\cite{ber81}.

\subsubsection{Deformation effects}

\begin{figure}[t]
\begin{center}
\includegraphics[scale=0.26]{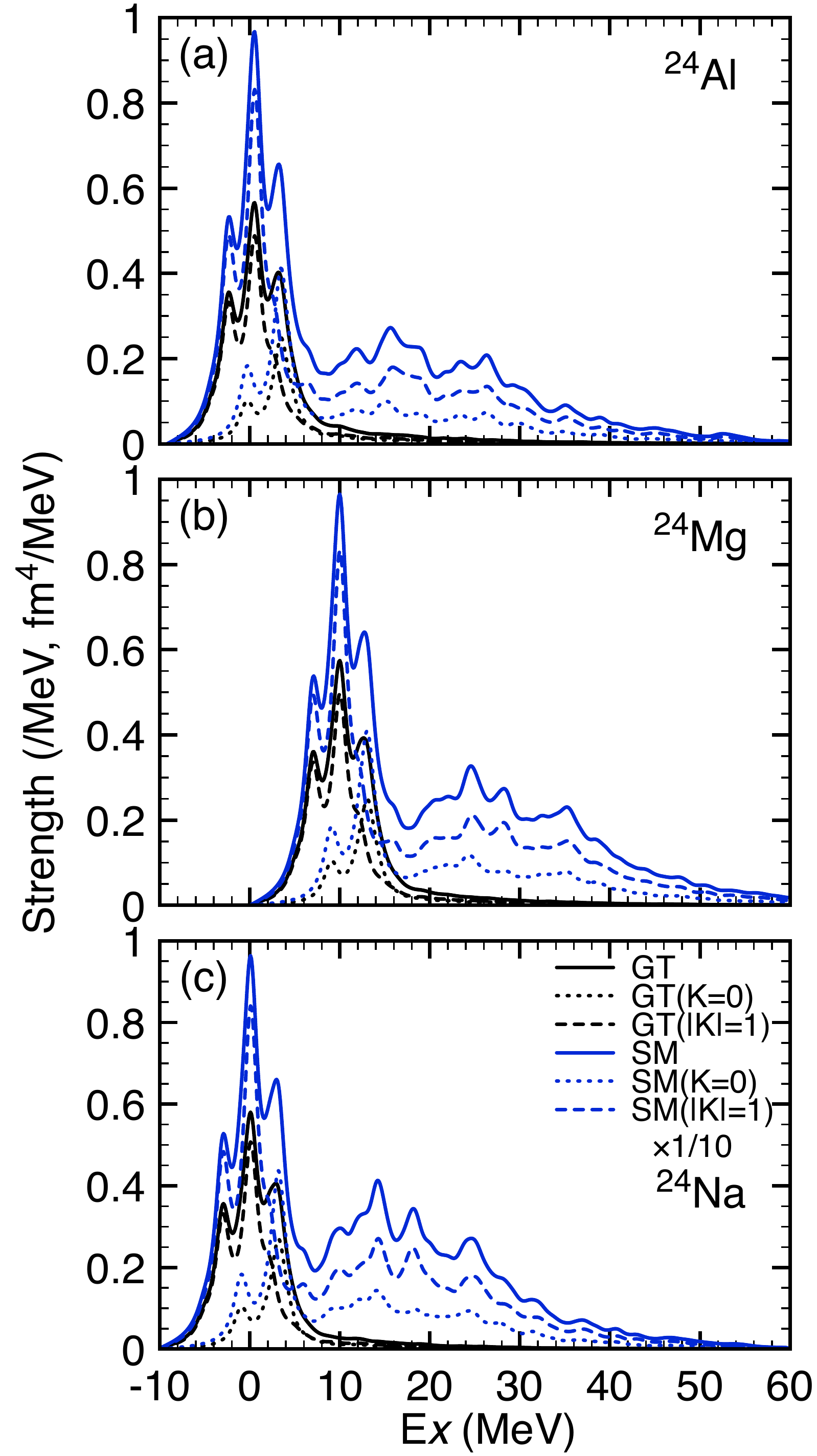}
\includegraphics[scale=0.26]{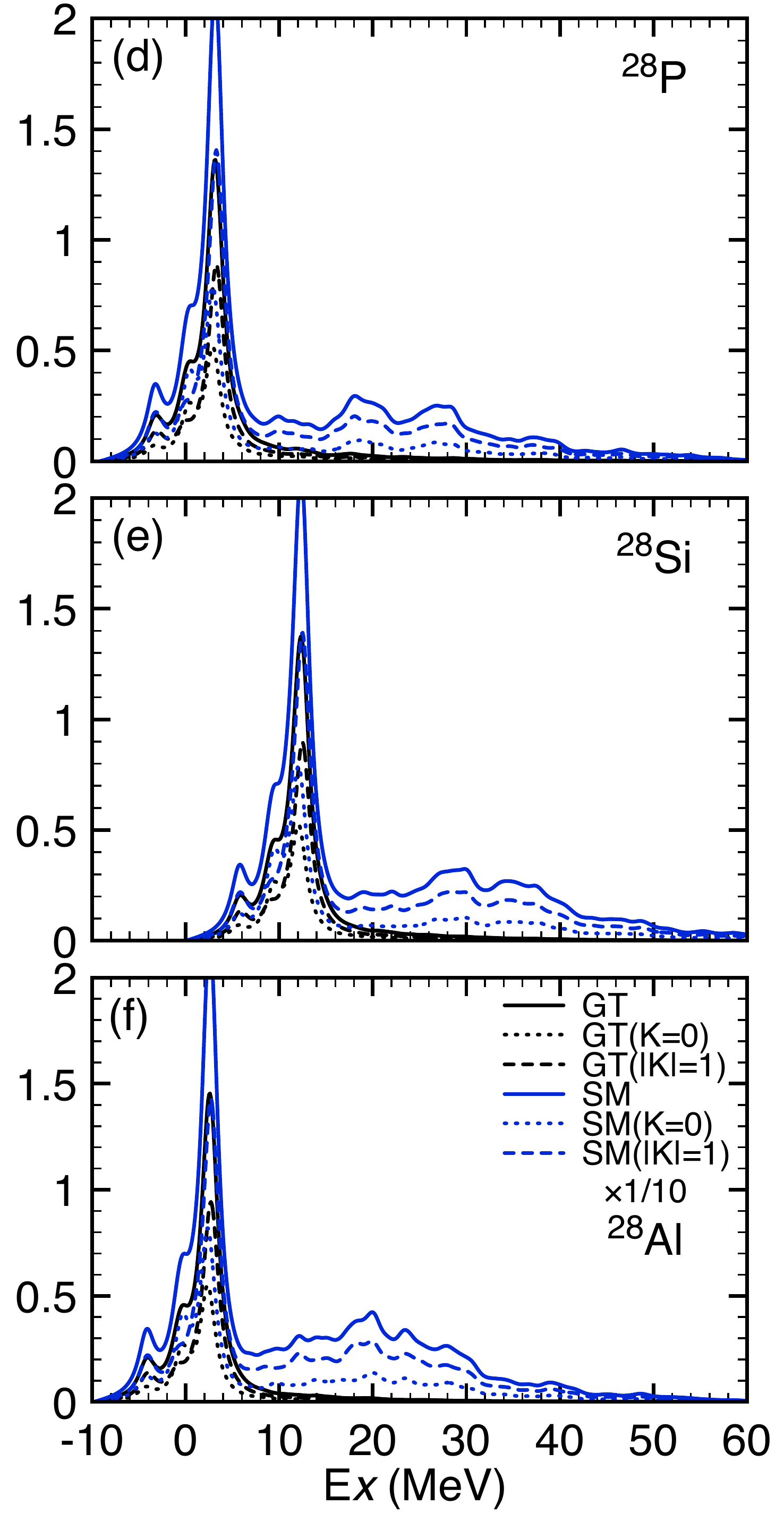}
\caption{\label{fig:mag_def} 
As Fig.~\ref{fig:ele_def} but for the Gamow--Teller and spin-monopole (SM) excitations. 
The strengths of $K=\pm 1$ are summed for $|K|=1$. The SM strengths are multiplied by $1/10$. 
The total strengths denoted by the solid lines include both the $J=1$ states with $K=0$ and those with $K=\pm 1$, 
while the dotted and dashed lines show the $K=0$ and $|K|=1$ states, respectively. 
}
\end{center}
\end{figure}

\begin{figure}[t]
\begin{center}
\includegraphics[scale=0.26]{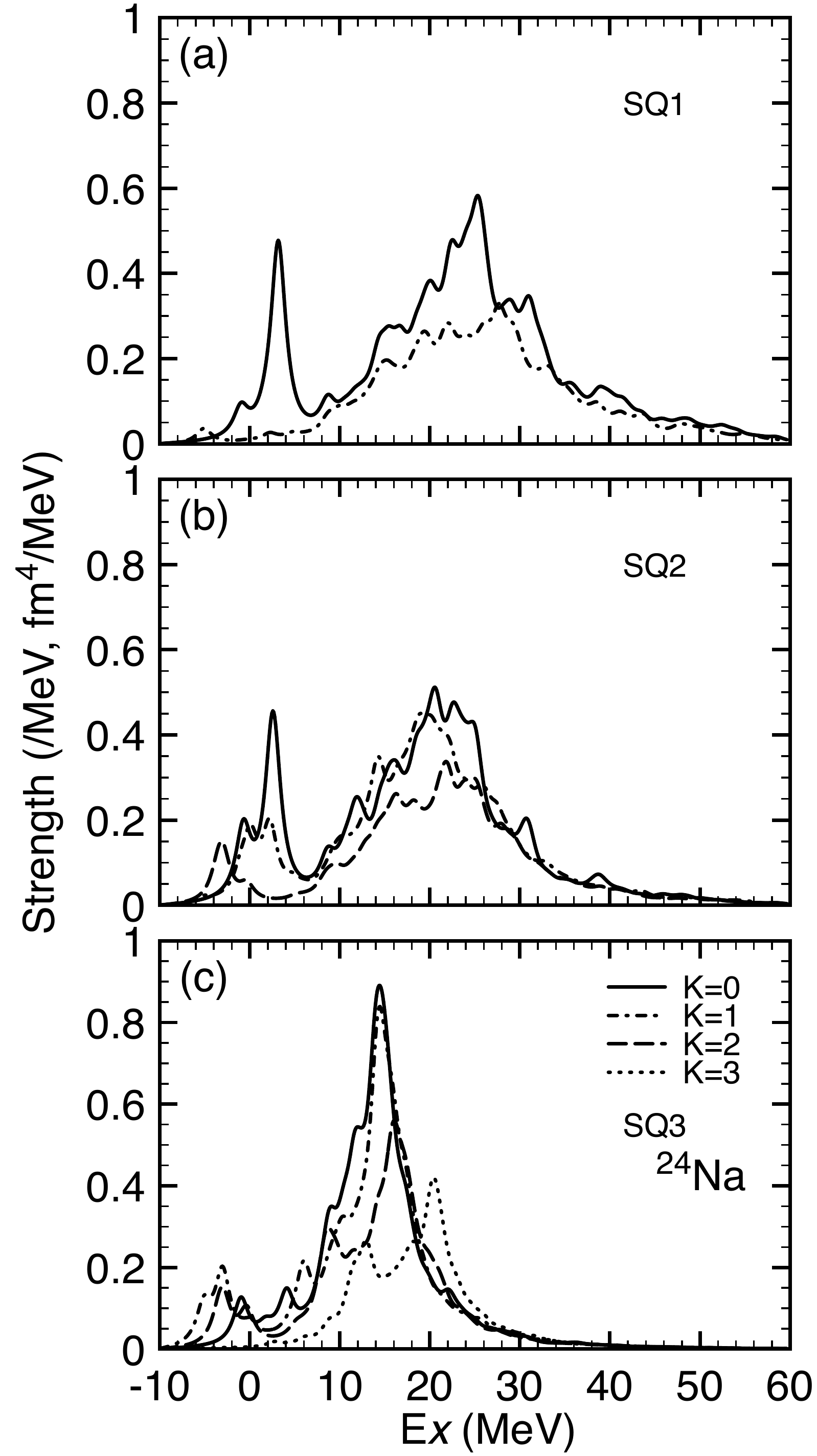}
\includegraphics[scale=0.26]{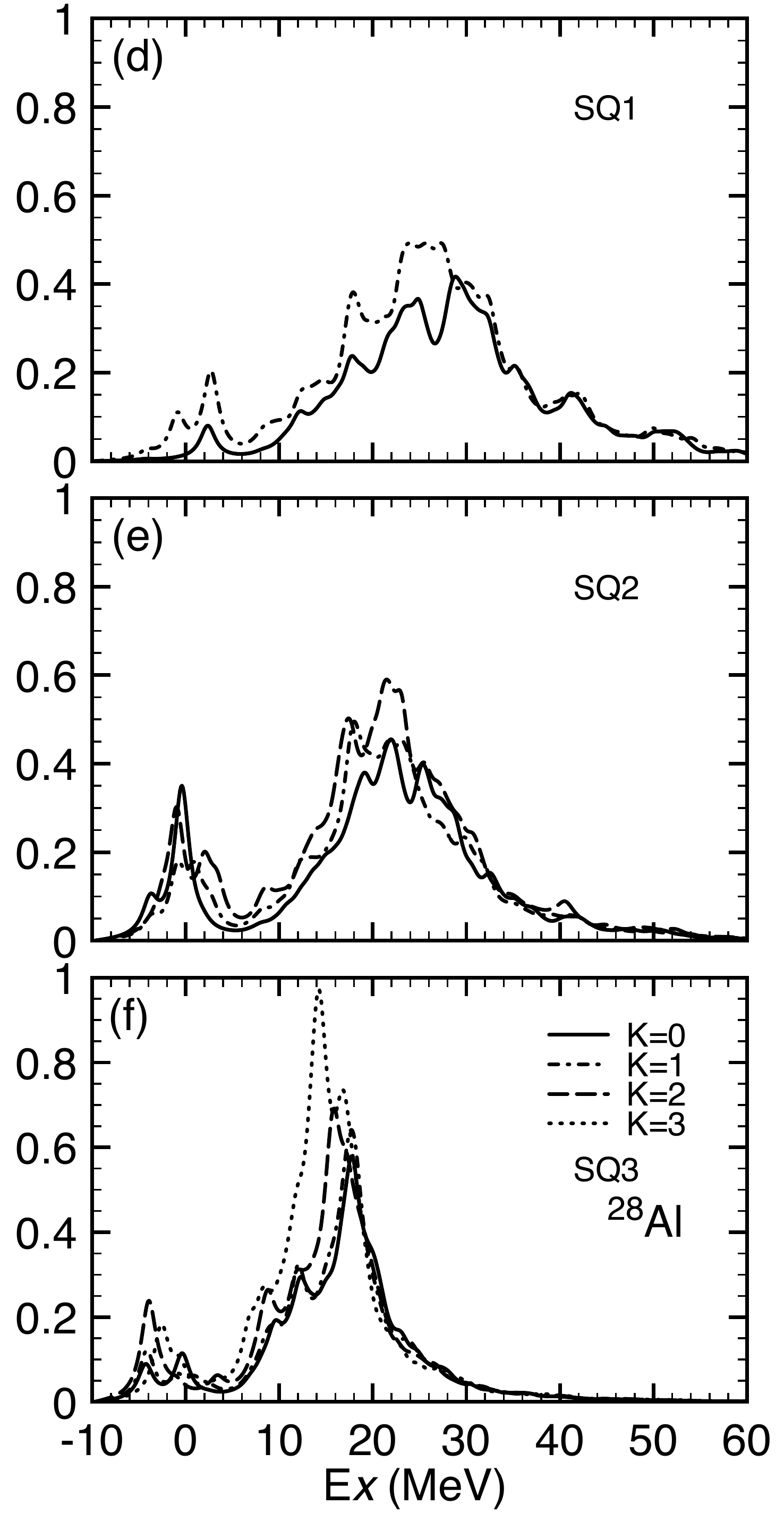}
\caption{\label{fig:mag2_def} 
As Fig.~\ref{fig:ele_def} but for the spin-quadrupole (SQ) excitations in the $\mu=+1$ channel. 
}
\end{center}
\end{figure}

I am going to investigate the deformation effects. 
Figure \ref{fig:mag_def} shows the GT and SM transition-strength distributions in $^{24}$Mg and $^{28}$Si. 
The total strengths denoted by the solid lines include the GT and SM transitions to both 
the $J=1$ states with $K=0$ and those with $K=\pm 1$, 
while the dotted and dashed lines depict the $K=0$ and $|K|=1$ states, respectively. 
A large fraction of the SM strengths is found in low energy, where the GTR shows up. 
This characteristic feature is found in spherical nuclei as well~\cite{aue84,ham00}. 
As in the electric cases, the transition strength distributions in the three channels are similar to each other. 
Furthermore, the SM transition strengths in the $\mu=+1$ channel are enhanced 
because the Coulomb potential slightly expands the proton distribution, which leads to the asymmetry even in the $N=Z$ nuclei. 

The strength distributions for $K=0$ and $K=1$ are different since the ground state is deformed. 
However, the $K$-splitting does not show a ``universal behavior'' that 
the $K=0$ states are shifted lower (higher) in energy in a prolately (oblately) deformed nucleus. 
This is because the GT operator does not change the spatial structure and the SM operator does not depend on the spatial direction. 
Furthermore, the ground state is time-even: $\langle \sigma_\nu  \rangle=0$. 
The $K$-splitting occurring in the GT and SM excitations are due not to the collective deformation 
but to the underlying shell structure. 
In $^{24}$Mg, the $K=1$ states appear lower in energy than the $K=0$ states, although the ground state is prolately deformed. 
The Fermi levels of neutrons and protons are both located in between the [211]3/2 and [202]5/2 orbitals. 
The $K=1$ state is mainly generated by the $[211]3/2 \to [202]5/2$ and $[211]3/2 \to [211]1/2$ excitations, 
while the $K=0$ state is constructed, {\it e.g.}, by the $[220]1/2 \to [211]1/2$ excitation, both of which are 
far from the Fermi level. Thus, the $K=1$ states appear lower in energy. 

I then investigate the SQ excitations. 
Since the SQ operator involves the spherical harmonics $Y_{2\nu}(\hat{r})$, 
the $K$-dependence can be attributed to nuclear deformation. 
However, the $K$ quantum number is composed of the $z$-component of angular momentum, reflecting the nuclear shape, and intrinsic spin, 
it is not apparant to expect a direct correspondence between the $K$-splitting and the nuclear deformation.

As discussed so far, the strength distributions in the $\mu=0$ and $\pm1$ channels are similar to each other for the $N=Z$ light nuclei. 
Thus, I show in Fig.~\ref{fig:mag2_def} the transition-strength distribution in the $\mu=+1$ channel only. 
One sees that the distributions for each $K$ are different. 
The $K=0$ ($K=1$) strengths are enhanced in a prolately (oblately) deformed nucleus for $J=1$. 
A universal feature of the $K$-splitting can be seen for $J=3$: 
the lower (higher)-$K$ states appear lower in energy and possess enhanced strengths in a prolately (oblately) deformed nucleus.
However, it is not easy to distinguish the strength distributions of each $K$ for $J=2$.

In the electric case, the $K$-dependence of the transition strengths was evaluated qualitatively by 
looking at the NEWSR values using Eq.~(\ref{eq:NEWSR_def}). 
The NEWSR values for the GT and SM excitations are essentially the same as those assuming spherical symmetry~(\ref{eq:NEWSR}): 
the spatial function $f(r)$ is constant for the GT operator, 
and that for the SM operator is $r^2$, which is scalar. 
However, one needs to consider the $K$-dependence for the SQ excitations. 
The NEWSR for the SQ excitations with $(J,K)$ reads
\begin{align}
&m_{L=2 (J,K)}^{-1}-m_{L=2 (J,K)}^{+1} 
\notag\\
&=\left\{
\begin{array}{ll}
\dfrac{1}{8\pi}(N\langle 4z^4 +\rho^4+5\rho^2 z^2\rangle_\nu - Z\langle \cdots \rangle_\pi) & (1,0) \\
\dfrac{1}{16\pi}(N\langle 2z^4 +5\rho^4+7\rho^2 z^2\rangle_\nu - Z\langle \cdots \rangle_\pi) & (1,1) \\
\dfrac{15}{8\pi}(N\langle \rho^2 z^2\rangle_\nu - Z\langle \cdots \rangle_\pi) & (2,0) \\
\dfrac{5}{16\pi}(N\langle 2z^4 + \rho^4 - \rho^2 z^2\rangle_\nu - Z\langle \cdots \rangle_\pi) & (2,1) \\
\dfrac{5}{16\pi}(N\langle \rho^4 + 2\rho^2 z^2\rangle_\nu - Z\langle \cdots \rangle_\pi) & (2,2) \\
\dfrac{1}{16\pi}(N\langle 12z^4 +3\rho^4 \rangle_\nu - Z\langle \cdots \rangle_\pi) & (3,0) \\
\dfrac{1}{32\pi}(N\langle 16z^4 +5\rho^4 +16\rho^2z^2 \rangle_\nu - Z\langle \cdots \rangle_\pi) & (3,1) \\
\dfrac{5}{32\pi}(N\langle \rho^4 +8\rho^2z^2 \rangle_\nu - Z\langle \cdots \rangle_\pi) & (3,2) \\
\dfrac{15}{32\pi}(N\langle \rho^4\rangle_\nu - Z\langle \cdots \rangle_\pi) & (3,3)
\end{array}
\right.,
\label{eq:NEWSR_def2}
\end{align}
where $\langle \cdots \rangle_\pi$ denotes the expectation value of the first term by replacing neutrons with protons. 
In deriving these sum rule values, I assume that $J^\pi$ of the ground state of the mother nucleus is $0^+$: 
the time-odd densities vanish in the ground state.   
For $J=1$, the $K=0$ ($K=1$) strengths are characterized by a large $\langle z^4\rangle$ ($\langle \rho^4\rangle$) term. 
Since the prolate (oblate) deformation produces the large $\langle z^4\rangle$ ($\langle \rho^4\rangle$) value, 
the above finding can be reasonably understood. 
The `stretched' $J=3$ excitation is relatively simple, particularly the $K=0$ and $K=3$ states.  
The prolate (oblate) deformation gives larger strengths in the $K=0$ ($K=3$) states. 
A similar feature has been found in the `stretched' $J=2$ spin-dipole excitation in deformed nuclei 
though the $K$-dependence is not clear for the $J=0$ and $J=1$ excitations~\cite{yos20}.

\begin{figure}[t]
\begin{center}
\includegraphics[scale=0.32]{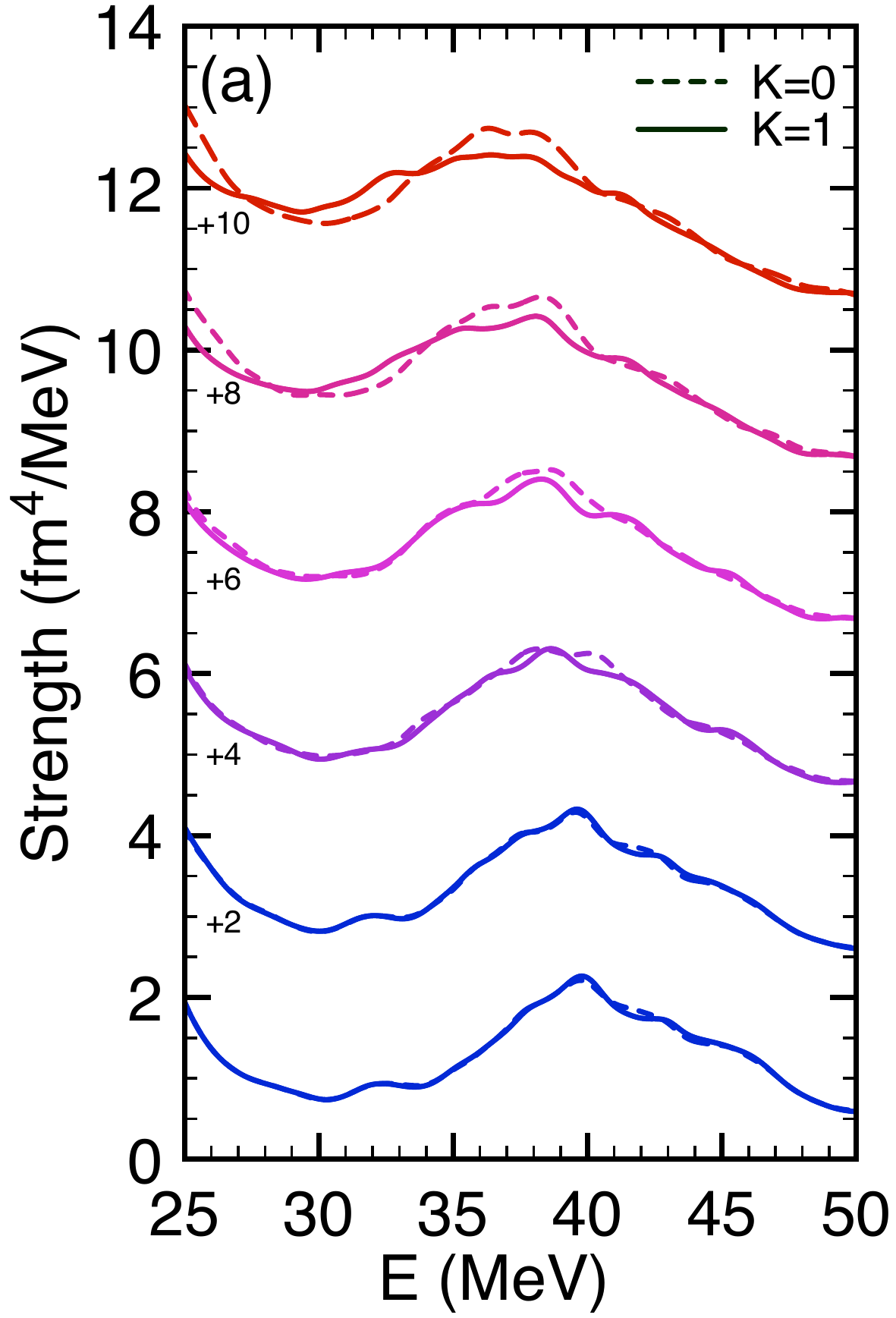}
\includegraphics[scale=0.32]{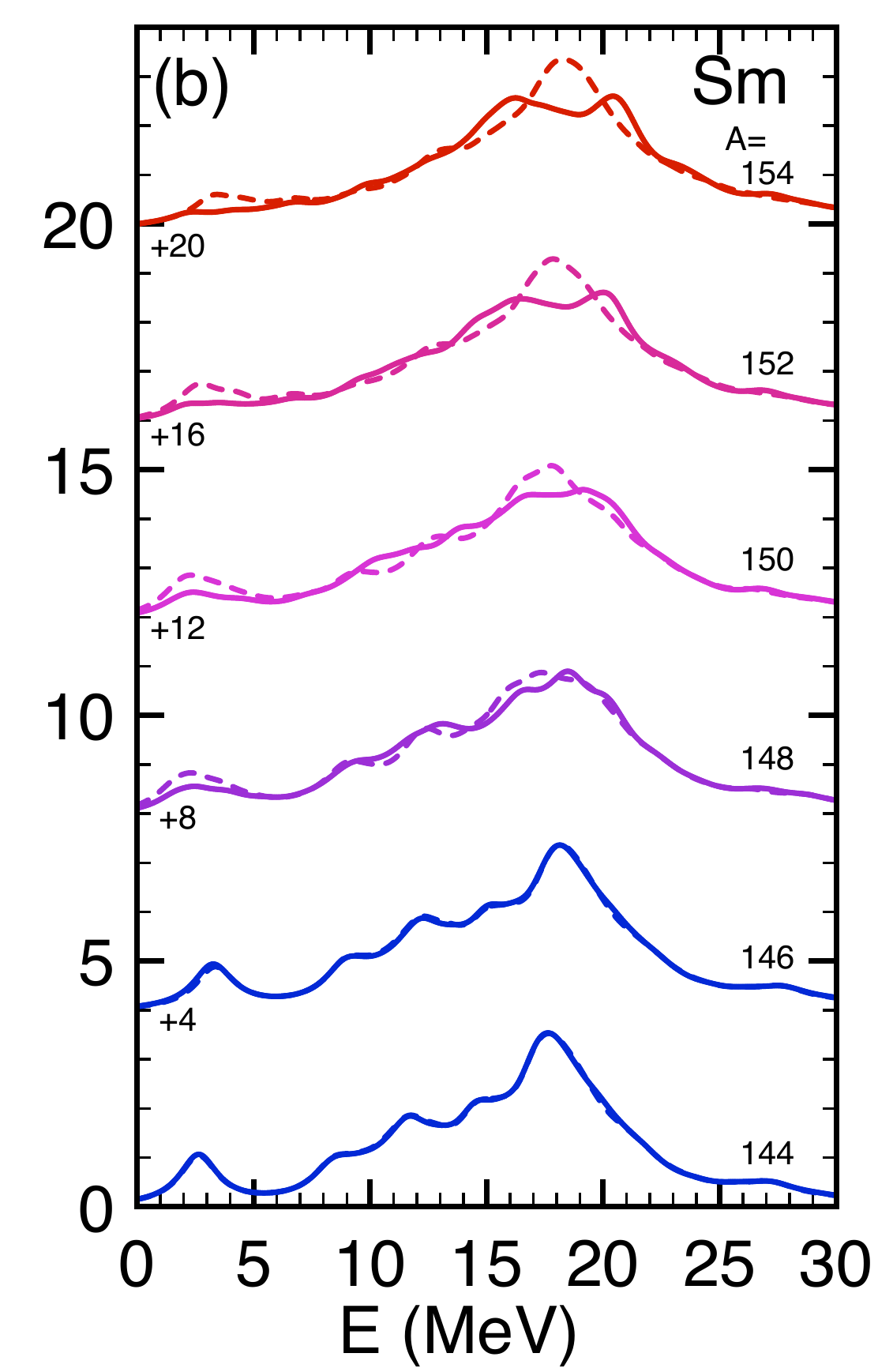}
\caption{\label{fig:Sm_SM} 
As Fig.~\ref{fig:Sm_mono} but for the spin-monopole (SM) excitations. The strengths for $K=0$ and $K=1$ are separately depicted by 
the dashed and solid lines, respectively. 
}
\end{center}
\end{figure}

According to the coupling between the monopole and the $K=0$ component of the quadrupole excitations seen in the electric case, 
which is universal both in the IS and IV excitations, 
one is tempted to expect the spitting of the SM strengths to appear due to 
coupling to the $K=0$ and $K=1$ components of the SQ excitations in deformed nuclei. 
Figure~\ref{fig:Sm_SM} shows the SM transition strengths in the Sm isotopes with $A=144\textendash154$. 
Here, the $K=0$ and $K=1$ strengths are displayed separately. 
It is hard to see the deformation effects in these distributions either in the $\mu=+1$ and $\mu=-1$ channels. 
One reason is that for the electric quadrupole excitations, the $K=0$ strengths are concentrated in a single peak; 
however, in the current case, the $K=0$ and $K=1$ strengths of the SQ excitations are widely spread out in 30--40 MeV depending on $J$. 
Another reason is that the SM strengths distribution is broadened irrespective of the nuclear shape. 

\section{Summary}\label{summary}

I have investigated the electric (non spin-flip) and magnetic (spin-flip) IV monopole and quadrupole modes of excitation. 
To obtain the generic features of the IV excitations, the neural $(\mu=0)$ and charge-exchange $(\mu=\pm1)$ channels have been considered simultaneously. 
Furthermore, I have explored open-shell nuclei to obtain unique features associated with nuclear deformation. 
To this end, I employed the nuclear energy-density functional (EDF) method: the Skyrme--Kohn--Sham--Bogoliubov 
and the quasiparticle random-phase approximation were used to describe the ground state and the transition to excited states.

A strong concentration of the monopole strengths in the energy region of the IAR has been found regardless of nuclear deformation. 
In addition, a resonance structure appears in high energy, which is generated mainly by particle--hole configurations with $2\hbar \omega_0$ excitation. 
The $K$-splitting occurs in the electric quadrupole excitations due to deformation. 
The lower (higher) $K$ states appear lower (higher) in energy in a prolately deformed nucleus: 
the opposite in an oblately deformed nucleus. 
Thus, the $K$-splitting of the GQR is universal in the IS and IV excitations. 
Furthermore, the coupling to the $K=0$ component of the GQR brings about the splitting of the monopole strengths in all the channels of IV excitation.

Similarly to the electric excitations, I have found 
a strong concentration of the spin-monopole strengths in the energy region of the GTR regardless of nuclear deformation. 
The $J=3$ states appear lowest in energy among the spin-quadrupole resonances. 
The $K$-splitting occurs in the spin-monopole and spin-quadrupole excitations. However, 
the relation between the energy-ordering depending on $K$ and the deformation is not apparant: 
the $K$ splitting in the spin-monopole excitation is due to the change in the underlying shell structure similarly to the GTR, 
and that in the $J=3$ spin-quadrupole resonance follows the universal trend.

\begin{acknowledgments} 
This work was supported by the JSPS KAKENHI (Grants No. JP19K03824 and No. JP19K03872). 
The numerical calculations were performed on Yukawa-21 
at the Yukawa Institute for Theoretical Physics, Kyoto University.

\end{acknowledgments}

\end{document}